\documentclass[aps,pra,twocolumn,groupedaddress,showpacs]{revtex4}
\usepackage{epsfig}
\usepackage{amsmath}
\usepackage{amsfonts}

\def\bE{\mathbf E}

\def\bx{\mathbf x}

\def\bk{\mathbf k}

\def\b0{\mathbf 0}

\begin{document}

\title{On the effective permittivity of finite inhomogeneous objects}
\author{Shreyas B. Raghunathan and Neil V. Budko}
\affiliation{Laboratory of Electromagnetic Research, Faculty of Electrical
Engineering, Mathematics and Computer Science,
Delft University of Technology,
Mekelweg 4, 2628 CD Delft, The Netherlands}
\email{n.v.budko@tudelft.nl}

\date{\today}

\begin{abstract}
A generalization of the S-parameter retrieval method for finite three-dimensional inhomogeneous
objects under arbitrary illumination and observation conditions is presented. 
The effective permittivity of such objects may be rigorously defined
as a solution of a nonlinear inverse scattering problem. In this respect the problems of 
S-parameter retrieval, effective medium theory, and even the derivation of the macroscopic 
electrodynamics itself, turn out to be all mathematically equivalent. We confirm analytically and 
observe numerically effects that were previously reported in the one-dimensional strongly inhomogeneous 
slabs: the non-uniqueness of the effective permittivity and its dependence on the illumination and 
observation conditions, and the geometry of the object. Moreover, we show that, although the 
S-parameter retrieval of the effective permittivity is scale-free at the level of problem statement, the exact solution of this 
problem either does not exist or is not unique. Using the results from the spectral analysis 
we describe the set of values of the effective permittivity for which the 
scattering problem is ill-posed. Unfortunately, real nonpositive values, important for negative
refraction and invisibility, belong to this set. We illustrate our conclusions using a numerical reduced-order inverse 
scattering algorithm specifically designed for the effective permittivity problem.
\end{abstract}


\maketitle

\section{Introduction}
The notion of an effective permittivity (conductivity, permeability) has been introduced within the effective-medium theory and 
was meant as a simplifying approximation of the scattering model for objects exhibiting inhomogeneity on scales much smaller 
than the scale of the spatial variation of the incident field \cite{Choy1999}. 
Thus, a gas, a colloid, a suspension, or a powder mixture could be modelled as
homogeneous media with some effective permittivities when interacting with a field having wavelength 
much larger than the size of the constituent particles and the average distance between them. 
Extension of the effective medium approximation beyond its natural
limits of applicability, i.e. for smaller wavelengths where the multiple interparticle scattering is more pronounced, 
leads to effective parameters exhibiting strange and sometimes exotic behavior. 
For example, if we insist on describing the Earth atmosphere as a homogeneous medium, then
the naive explanation of the sky's blue color could be a particular frequency-dependence of this homogeneous effective permittivity, 
so that it filters out other frequencies of the visual spectrum. In reality, as we know, the opposite happens. The 
high-frequency blue light is Rayleigh-scattered by the particles (molecules) of the atmosphere, so that even 
the parts away from the direct optical path to the Sun shine with blue light that reaches our eyes.
Whereas the low-frequency red light passes the atmosphere almost without scattering and causes the
red color of the sunset. This illustrates another important property of all effective scattering models. 
Should we decide to view the sky as a homogeneous medium, then we would have to introduce at least two 
effective models -- one for the day and another for the evening, i.e. our effective model would depend on the 
illumination and observation conditions. It was also noticed in \cite{Bohren1985} that the
effective model of an electrically inhomogeneous atmosphere may show non-zero magnetic contrast
-- a conclusion of great importance to the present studies of composite media.

In recent years the effective medium approach has been extensively used beyond its natural applicability
limits, especially for modelling the response of metamaterials and photonic crystals. Moreover, the effective parameters of 
such composite media seem nowadays to carry more physical meaning than a mere simplifying approximation 
of the scattering model. For example, the following practical question may be asked: can a composite medium 
be used in the same way as one would use, say, a borosilicate glass, to cut out a lens or a light-bending 
coating (invisibility cloak)? Or is there something about strongly inhomogeneous composites that makes them different from ordinary 
dense media in this respect? It should be mentioned, of course, that the very notions of the macroscopic permittivity and permeability 
of those ``ordinary'' media are themselves the product of the so-called continuum approximation, which is essentially 
an effective medium theory applied within its ``natural'' limits. Hence, the problem is, in fact, older and more fundamental.
It concerns the applicability, accuracy, and the physical meaning of the effective modelling as such. 

Partly due to the popularity of metamaterials, the derivation techniques and even the very concept of effective 
permittivity (permeability) have been recently revisited by many authors [3--32]. There are two main ways in which 
one can derive the effective permittivity of an inhomogeneous object. One is the traditional approach
of macroscopic electrodynamics known as homogenization. It attempts to derive a local constitutive
relation between the averaged field and a simplified permittivity (conductivity, permeability) function,
e.g. averaged over a representative cell or simply constant. Application of the homogenization to metamaterials is reviewed
in \cite{Smith2006, Menzel2009}. The authors recognize the problem, which the relatively large and strongly scattering metamaterial
particles pose for the applicability of this traditional effective medium approach and view homogenization 
as a complementary method to the so-called S-parameter retrieval technique. The latter technique, 
whose application to metamaterials was pioneered in \cite{Smith2002}, is the other popular way 
of looking at the problem where instead of explicitly 
deriving a local constitutive relation one is simply matching the observed (simulated or measured) field from 
a composite inhomogeneous slab to the field from a homogeneous slab of the same thickness. 
Often this is also the way the exotic properties of metamaterials and photonic slabs are verified experimentally. 
From the theoretical point of view the S-parameter retrieval method is very attractive, 
because it seems to be immune against the aforementioned 
scalability problems inherent to homogenization techniques. Indeed, one does not care what the relative scale 
of inhomogeneities is, as long as there is a match with a field from a homogeneous object of the same outer shape.
There exists a third method, which includes the measured internal field in the matching procedure \cite{Popa2005}. 
However, as we show here, it is essentially an extension of the S-parameter retrieval technique.

Despite its generality, the S-parameter retrieval poses other questions. First of all, the retrieved effective
parameters for a homogeneous slab turn out to be non-unique \cite{Smith2002, Seetharamdoo2005}. 
Secondly, their values are sensitive to the location of the slab boundary \cite{Smith2002}, orientation and
regularity of the cells \cite{Wang2006},  and depend on the angle of incidence of the illuminating plane 
wave \cite{Smith2005}. This dependence on the wavevector $\bk$ is considered by some to be the sign
of anisotropy and spatial dispersion \cite{Cabuz2008} and has lead others \cite{Menzel2008a, Menzel2008b, Menzel2009} to question
the usefulness of the effective medium parameters as such, since they do not serve their original purpose of 
simplifying the propagation model any more. Also, the complex effective permittivity may sometimes 
show a negative loss, i.e. gain, \cite{Seetharamdoo2005} and larger than expected (from homogenization) 
positive losses \cite{Seetharamdoo2005, Pimenov2006}. 

The theoretical analysis of the S-parameter retrieval method has been limited so far to the slab-like configurations where
an analytical solution for a homogeneous case is readily available.
Here we consider a generalization of this approach to finite three-dimensional structures, inhomogeneous, 
and not necessarily metamaterials.  
We apply the volume integral formulation and show that 
we deal with a special kind of inverse scattering problem. We demonstrate the mathematical equivalence of the generalized
S-parameter retrieval (or effective inversion) problem to the original problem of the effective medium theory.
Despite the lack of explicit analytical solutions in 3D, a combination of spectral analysis and inverse scattering 
theory confirms all the anomalous features of the S-parameter retrieval method in the present general 3D case 
and sheds new light on their mathematical origins. 

The most important and somewhat surprising results of our study 
are: in general, an exact effective model of lower complexity does not exist; if an effective permittivity exists, it is not unique,
and there is often no way one can choose a particular value on ``physical'' grounds; 
the goals of having an accurate effective scattering model and a unique effective permittivity contradict each other;
the effective model is singular for (infinitely) many real nonnegative values of permittivity (at least for electrically small objects). 
In most of the paper we limit ourselves to the retrieval of an effective permittivity of a homogeneous 
(lower complexity) effective model, although, effective models 
of higher complexity (anisotropic, magnetic) are briefly discussed as well. We illustrate the key effects on a number of numerical 
examples using an algorithm specifically designed for the problem. This algorithm is of interest in itself 
as it eases the computational burden of having to solve many forward scattering problems 
(one for each trial value of the effective permittivity) by employing the shift invariance of the Arnoldi iterative scheme.

\section{Analysis of the forward scattering problem}

\subsection{The volume integral equation method}
The most general definition of an effective scatterer is, in fact, quite simple and can be formulated as follows.
An effective scatterer has the same or similar outer shape as the original object; different, possibly simpler, internal structure; 
and produces the same (or approximately the same) scattered field as the original 
under identical illumination and observation conditions.
In the following section we shall make this definition more precise. 
For the moment, let us consider a general inhomogeneous, isotropic, finite object occupying the spatial domain $D$.
Its constitutive parameters are the spatially varying complex, possibly dispersive, dielectric 
permittivity $\varepsilon(\bx,\omega)$ and constant magnetic permeability $\mu_{0}$. The surrounding medium
is infinite and homogeneous with the vacuum parameters $\varepsilon_{0}$ and $\mu_{0}$. Thus, the object has
no magnetic contrast with respect to the background, whereas its electric contrast is given by the function
 \begin{align}
 \label{eq:ChiE}
 \chi(\bx,\omega)=\frac{\varepsilon(\bx,\omega)}{\varepsilon_{0}}-1.
 \end{align}
Obviously, $\chi(\bx,\omega)=0$, if $\bx\notin D$. Let the object be illuminated by an external source, which
generates the known electric field $\bE^{\rm in}(\bx,\omega)$ in vacuum. The total electric field $\bE(\bx,\omega)$ inside 
the scatterer satisfies the following volume integral equation:
 \begin{align}
 \label{eq:IntegralEquation}
 \begin{split}
 \bE(\bx,\omega)-\int\limits_{\bx'\in D}{\mathbb G}(\bx-\bx',\omega)&\chi(\bx',\omega)\bE(\bx',\omega)\,{\rm d}\bx'
 \\ 
 &=
 \bE^{\rm in}(\bx,\omega), \;\;\;\bx\in D,
 \end{split} 
 \end{align}
where ${\mathbb G}(\bx,\omega)$ denotes the Green's tensor. This is a simplified notation, of course. The actual form of
the integral operator can be found in \cite{Rahola2000, Samokhin2001} or within the literature on the Discrete Dipole 
Method \cite{Yurkin2007}. Some theoretical results of importance to our discussion have been accumulated over the years.
In this section we briefly review these results 
with the emphasis on understanding and physical meaning rather than mathematical rigour.

\subsection{Existence and uniqueness}
In operator notation equation (\ref{eq:IntegralEquation}) can be written as
\begin{align}
\label{eq:IEop}
[I-GX]u=u^{\rm in},
\end{align}
where $I$ is the identity operator, $G$ is the integral operator with the Green's tensor kernel, $X$ is a ``diagonal''
operator of pointwise multiplication with the contrast function, $u$ is the unknown total field, and $u^{\rm in}$ is the known 
incident field. In our formulation the linear operator $I-GX$ belongs to the class
of singular integral operators \cite{Mikhlin1980} and its kernel is {\it strongly} singular as opposed to the {\it weakly} 
singular kernels of the corresponding integral operators in one- and two-dimensional scattering problems. 
In \cite{Samokhin2001} the mathematical equivalence of the integral equation (\ref{eq:IntegralEquation}) 
to the frequency-domain Maxwell's 
equations with the radiation boundary condition was shown (for H{\"o}lder-continuous incident fields), 
and the necessary and sufficient condition was obtained for the existence of a solution. 
In the isotropic case with H{\"o}lder-continuous contrast functions the solution of (\ref{eq:IEop}) 
exists if and only if 
 \begin{align}
 \label{eq:ExistenceCond}
 \varepsilon(\bx,\omega)\ne 0, \;\;\;\;\bx\in{\mathbb R}^{3}.
 \end{align}
Two sufficient conditions that guarantee the uniqueness of the solution are known. 
One is the presence of non-zero losses, i.e. ${\rm Im}\,\varepsilon(\bx,\omega)>0$ \cite{Colton1992b, Samokhin2001}. 
In the lossless case, ${\rm Im}\,\varepsilon(\bx,\omega)=0$, a three times continuously differentiable permittivity function (with respect to all 
coordinates and in ${\mathbb R}^{3}$) is also sufficient  for the uniqueness \cite{Samokhin2001}. 

Some authors prefer to work with an integro-differential form of equation (\ref{eq:IntegralEquation}),
where the kernel of the integral operator is kept weakly singular (three-dimensional scalar Green's function
of the Helmholtz equation), and the two spatial derivatives (grad-div operator) are kept outside the 
integral \cite{Colton1992a, Zwamborn1992}.
Although, the analysis is more complicated in that case and involves Sobolev rather than Hilbert spaces, 
in \cite{VanBeurden2003} a condition similar to the existence condition (\ref{eq:ExistenceCond}) was also obtained.

\subsection{Spectrum}
One of the advantages of considering the operator $I-GX$ in its strongly singular form
is that it naturally acts on the Hilbert space and one can apply a well established theory \cite{Mikhlin1980} 
to analyze its spectrum. The physical importance of the eigenfunctions and eigenvalues of $I-GX$ stems from the fact 
that they describe the {\it spatial} spectrum of the field, similar to the eigenmodes of a closed resonator or the plane 
waves in the one-dimensional case. Detailed analysis of this problem was presented in \cite{Budko2006a}, where the spectrum 
was found to contain not only the usual eigenvalues, but a nontrivial essential part as well. This is also a purely 
three-dimensional phenomenon, related to the strong singularity of the kernel and not present in one- and 
two-dimensional scattering \cite{Kleinman1990}. 
The difference between the eigenvalues and the essential spectrum can be explained as follows. 
Eigenvalues $\lambda$ and eigenfunctions $u_{\lambda}$ satisfy
 \begin{align}
 \label{eq:Eigenvalues}
 [I-GX]u_{\lambda}=\lambda u_{\lambda},
 \end{align}
where $u_{\lambda}$ belongs to the functional space in question -- Hilbert space here. That is to say
that the eigenfunctions may be viewed as some well-defined and, in principle, realizable spatial 
distributions of the field on $D$ -- an equivalent of the resonator modes. 
The exact location of eigenvalues in the complex plane is not known, only a bound is available:
 \begin{align}
 \label{eq:EigenvalueBound}
 \begin{split}
 {\rm Im}\varepsilon(\bx,\omega)
 -&{\rm Im}\varepsilon(\bx,\omega){\rm Re}\lambda+
 \\
 &\left({\rm Re}\varepsilon(\bx,\omega)-\varepsilon_{0}\right){\rm Im}\lambda\le 0,
 \\
 &\vert\lambda\vert<\infty.
 \end{split}
 \end{align}
The last inequality follows from the boundedness of the operator. 
The distribution of eigenvalues inside this wedge-shaped bound depends on the 
scatterer and applied frequency $\omega$. 
The essential spectrum, on the other hand, 
satisfies the Weyl definition:
 \begin{align}
 \label{eq:Essential}
 \begin{split}
 \lim\limits_{n\rightarrow\infty}
 &\left\Vert [I-GX]\Psi_{n} - \lambda_{\rm ess}\Psi_{n} \right\Vert =0,
 \\
 &\Vert\Psi_{n}\Vert=1,
 \end{split} 
 \end{align} 
where, while each $\Psi_{n}$ belongs to the Hilbert space, the sequence $\{\Psi_{n}\}$ does not have a convergent 
subsequence, and does not converge to any function in the usual meaning of the word. 
Thus, the ``essential'' mode does not represent any physically realizable 
field distribution. It was shown in \cite{Budko2006a} that the essential spectrum of $I-GX$ is
given explicitly as
 \begin{align}
 \label{eq:EssentialSpectrum}
 \lambda_{\rm ess}=\frac{\varepsilon(\bx,\omega)}{\varepsilon_{0}}, \;\;\;\;\bx\in{\mathbb R}^{3},
 \end{align}
i.e. it contains all values of the relative permittivity. 
While eigenvalues are ``discrete'', i.e. isolated points in the complex plane,
the essential spectrum is obviously a dense set, if $\varepsilon(\bx,\omega)$ is 
H{\"o}lder-continuous as presumed in the analysis. It was also shown in \cite{Budko2007} that the
corresponding sequence $\{\Psi_{n}\}$ is, in fact, a mollifier of the square root of the three-dimensional
Dirac delta-function -- a very exotic distribution localized around the position $\bx_{\rm c}$ 
determined by the corresponding point of the essential spectrum 
$\lambda_{\rm ess}=\varepsilon(\bx_{\rm c},\omega)/\varepsilon_{0}$. 
It is clear that definition (\ref{eq:Essential}) encompasses eigenvalues as well, since
with eigenvalues the sequence $\{\Psi_{n}\}$ simply converges to a function 
$u_{\lambda}$ from (\ref{eq:Eigenvalues}), which does belong to the Hilbert space. 

A connection of the essential spectrum with physics was suggested in \cite{Budko2007} via the notion of the
pseudospectrum and its pseudomodes \cite{Trefethen2005}. If the eigenvalues of $I-GX$ are 
defined as complex numbers $\lambda$ for which
 \begin{align}
 \label{eq:EigenvaluesDef}
 \left\Vert\left[I-GX-\lambda I\right]^{-1}\right\Vert=\infty,
 \end{align}
then the $\epsilon$-pseudospectrum of $I-GX$ may be defined as a set of all $\lambda_{\rm ps}$ satisfying:
 \begin{align}
 \label{eq:PseudoSpectrum}
 \left\Vert\left[I-GX-\lambda_{\rm ps} I\right]^{-1}\right\Vert>\frac{1}{\epsilon},
 \end{align}
for some $\epsilon>0$. By analogy, we extend the Weyl definition (\ref{eq:Essential}) as 
 \begin{align}
 \label{eq:EssentialPseudo}
 \lim\limits_{n\rightarrow N(\epsilon)}
 \left\Vert [I-GX]\Psi_{n} - \lambda_{\rm ps}\Psi_{n} \right\Vert \le \epsilon.
 \end{align} 
In this case, if $N(\epsilon)<\infty$, the sequence $\{\Psi_{n}\}$ will stop
at some highly localized function $\Psi_{N}$, still in the Hilbert space and a physically realizable 
field distribution. It is a pseudomode though, since it ceases to be a function for 
$\epsilon\rightarrow 0$ and $N\rightarrow\infty$.

Both the eigenvalues and the essential spectrum play important roles in resonant phenomena \cite{Budko2006b}.
If either an eigenvalue or the point of essential spectrum gets close to the zero 
of the complex plane, then a resonance is observed and most of the electromagnetic energy
will be accumulated in the corresponding eigenfunction or a pseudomode, which therefore
will determine the spatial distribution of the total field on $D$. 
Since all eigenfunctions and pseudomodes 
are rapidly decaying, if continued outside $D$, the resonances will generally lead to an increase of the field strength
inside the scatterer and a decreased field outside $D$ -- something one expects 
and observes. The difference between the eigenvalue-based and the essential-spectrum-based resonances is in their physical 
origins. The distribution of eigenvalues is strongly influenced by the geometry of the scatterer
and operating frequency. To get an eigenvalue close to zero one needs a proper
combination of the size and frequency, similarly to a half-wavelength condition in a 
dipole antenna. Whereas, to get the essential spectrum close to zero we only need to have 
the permittivity of the object close to zero at some arbitrary point in $D$, independently
of the object size and geometry. This is the main difference
between a material-based (microscopic) and a geometry-based resonances. A natural microscopic
resonance occurs if the permittivity exhibits anomalous dispersion.
Since from the outside both resonances may look the same (e.g. decreased transmission or reflection), 
we can, indeed, view a metamaterial or a photonic crystal as a composite scatterer which mimics 
a natural microscopic resonance by a geometry-based one.

Apart from describing the dominant field distributions during resonances, the use of eigenfunctions 
is rather limited. This has to do with the fact that they are rarely given explicitly, but also with 
another unfortunate property of the 
operator $I-GX$, its non-normality. Recall that an operator is non-normal if it does not commute 
with its own adjoint, i.e., $[I-GX]^{*}[I-GX]\ne[I-GX][I-GX]^{*}$. Non-normal operators are not
unitary diagonalizable, hence, they cannot be diagonalized using their eigenfunctions. This
is not specific to three dimensions and seems to be a fundamental property of the frequency-domain 
electromagnetic scattering, with its true physical significance yet to be uncovered.  

\subsection{The scattered field}
When nondestructive measurements are performed, one is usually able to measure the field scattered 
outside $D$ only. The scattered field is defined as $\bE^{\rm sc}=\bE-\bE^{\rm in}$.
Once the total field $\bE(\bx,\omega)$ on $D$ is obtained by solving the integral equation (\ref{eq:IntegralEquation}),
the scattered field on some measurement domain $S$ is obtained by simply evaluating the integral:
 \begin{align}
 \label{eq:ScatteredField}
 \begin{split}
 \bE^{\rm sc}(\bx,\omega)=\int\limits_{\bx'\in D}{\mathbb G}(\bx-\bx',\omega)&\chi(\bx',\omega)\bE(\bx',\omega)\,{\rm d}\bx',
 \\ 
 \bx\in S.
 \end{split} 
 \end{align}
In operator notation we shall write
 \begin{align}
 \label{eq:SCop}
 RXu=u^{\rm sc}.
 \end{align}
The integral operator $R$ is not singular, not even weakly, since $\bx\ne\bx'$ in the argument
of the Green's tensor. Hence, we have an integral operator with a smooth, absolutely integrable
kernel. It maps between the Hilbert space of functions with spatial support on $D$ -- object domain -- and the
Hilbert space of functions with spatial support on $S$ -- data domain. It is a compact operator
and as such is very different from the previously considered operator $I-GX$. 
The properties of $R$ are important in inverse scattering and have been discussed
in the corresponding literature \cite{Colton1992b, Dudley2000}. 
Compact operators have zero as the only accumulation point of their eigenvalues and
often have zero as an eigenvalue too. For example,
consider a single data-point so that $u^{\rm sc}$ is just a complex number
representing the measured complex amplitude of a single Cartesian component of the scattered 
field. Then, the integral formula
(\ref{eq:ScatteredField})-(\ref{eq:SCop}) can be written as an inner product
of two vector-valued functions:
 \begin{align}
 \label{eq:SingleData}
 u^{\rm sc}=\langle Xu,r\rangle = \langle w,r\rangle,
 \end{align}
where $r$ is a vector-valued function representing the complex conjugate of the $k$-th row of the Green's tensor $G_{km}(\bx-\bx',\omega)$ 
with fixed $\bx\in S$ and $\bx'$ taking values in $D$. We see that any $w_{N}\ne 0$ orthogonal to $r$ in the sense
of the inner product would produce zero scattered field. Analogously, if we measure the
scattered field at any finite number of points, and the operator $R$ represents
the so-called semi-discrete mapping, then it is always possible to find functions $w_{N}\ne 0$,
such that $Rw_{N}=0$. Such functions, often called non-radiating sources, belong 
to the null-space of operator $R$, denoted as ${\mathcal N}(R)$. The consequence of all this is that the
operator $R$ does not have a bounded inverse, and solution of the so-called
inverse-source problem $Rw=u^{\rm sc}$ with respect to $w$ may be not unique. 

In \cite{Popa2005} the measurements were reported {\it inside} a metamaterial
and the fields were compared (matched) not only on $S$, but on $D$ as well. Obviously,
however, no measurements are possible all over $D$ without destroying the
metal particles or perturbing the field distribution on their surfaces. 
Therefore, parts of $D$ will remain inaccessible, and the null-space of $R$ 
will, in general, remain nontrivial. On the other hand, the case where instead of actual 
measurements one simulates numerically the fields on $D$, corresponds to $RX=GX$, and 
may constitute a complete data-set discussed in the following section.

\section{Analysis of the inverse scattering problem}

\subsection{Effective inversion}
Let there be an inhomogeneous object with permittivity $\varepsilon(\bx,\omega)$, 
occupying the spatial domain $D$, illuminated by the incident field $u^{\rm in}$,
and producing the scattered field $u^{\rm sc}$ over some data domain $S$.
Naturally, we shall assume that for this real-world object there exists a unique 
solution of the forward scattering problem (\ref{eq:IEop}). Formally this solution
may be written as
 \begin{align}
 \label{eq:IEopSolution}
 u=[I-GX]^{-1}u^{\rm in}.
 \end{align}  
Hence, the scattered field is obtained as
 \begin{align}
 \label{eq:SCopSolution}
 u^{\rm sc}=RX[I-GX]^{-1}u^{\rm in}.
 \end{align}
This is the main equation of the inverse scattering theory, where one
wants to find the constitutive parameters of the object -- the diagonal operator $X$ --
from the knowledge of the incident and scattered fields. We can see from (\ref{eq:SCopSolution})
that the inverse scattering problem is not only nonlinear but also almost certainly ill-posed, due to the 
presence of the operator $R$. 

Our goal is to find another object, an effective scatterer, occupying the 
same spatial domain $D$, but having a different 
permittivity function $\varepsilon_{\rm ef}(\bx,\omega)\ne \varepsilon(\bx,\omega)$, 
such that the application of the incident field $u^{\rm in}$ will produce the same scattered field 
$u^{\rm sc}$ as the original object.
Hence, what we want is, in fact,
 \begin{align}
 \label{eq:EffInv}
 RX_{\rm ef}[I-GX_{\rm ef}]^{-1}u^{\rm in}=RX[I-GX]^{-1}u^{\rm in}.
 \end{align}
The effective permittivity function can be simpler than the original, 
e.g. homogeneous (constant) over $D$. It can also be the result of spatial averaging
or smoothing described mathematically as an application of some linear integral 
operator $A$ on the original permittivity function, i.e. $X_{\rm ef}=AE-I=E_{\rm ef}-I$,
where $E$ is the diagonal operator of relative permittivity. The only thing which really 
matters here is that $E_{\rm ef}\ne E$ and $X_{\rm ef}\ne X$. Hence, we would like the inverse scattering problem
(\ref{eq:SCopSolution}) to have at least and at most two different solutions, $X$ and 
$X_{\rm ef}$. 

This problem was considered in \cite{Budko1998, Budko1999} in a completely different context --
as a fast algorithm to determine the effective permittivity of a buried object with the goal to
discriminate landmines from stones and other targets. To distinguish (\ref{eq:EffInv})
from the standard inverse scattering problem (\ref{eq:SCopSolution}), the former is called the {\it effective 
inversion} problem. Obviously, it represents a generalization of the S-parameter retrieval
method. 

It is possible to reformulate the standard effective medium theory (EMT) along 
the same lines. In EMT and in the derivation of macroscopic permittivity  
an averaged, smoothed total field inside the object is introduced. Let us denote this procedure as $Bu$, so that the averaging 
of the field inside the original highly inhomogeneous scatterer, characterized by $X$, is obtained as
 \begin{align}
 \label{eq:AveTot}
 Bu=B[I-GX]^{-1}u^{\rm in}.
 \end{align}  
The main conjecture of the EMT and macroscopic electrodynamics is that the same field will be observed inside a suitably averaged
effective object $X_{\rm ef}=AE-I$, i.e.
 \begin{align}
 \label{eq:Homogenization}
 [I-GX_{\rm ef}]^{-1}u^{\rm in}=B[I-GX]^{-1}u^{\rm in}.
 \end{align}  
Since the field-averaging operator $B$ is also a compact integral operator with a smooth kernel,
the similarity with (\ref{eq:EffInv}) is obvious. The two problems become  
equivalent, if we $B$-average the left-hand side of (\ref{eq:Homogenization}) as well.
Hence, the fundamental problems of effective inversion, S-parameter retrieval method, 
effective medium theory, and macroscopic electrodynamics are all mathematically equivalent
up to the actual form of compact operators $A$, $B$, and $RX$, if both the true and
the effective fields are averaged in the latter two approaches. From now on
we shall concentrate on the S-parameter approach (effective inversion) as the only one where
the $RX$-operator is explicit.

\subsection{Non-existence} 
The first question we should ask ourselves is: does an effective scatterer 
exist at all?
To this end we recall a well-known property of inverse scattering problems.
While the compact operator $R$ does not have a bounded inverse, 
the inverse scattering problem may still have a unique, though unstable, solution \cite{Colton1992a}. 
The known uniqueness conditions in inverse scattering theory are sufficient, not necessary, 
and usually describe a particular set of incident fields
and a set of measured scattered fields which together guarantee that there is only one solution 
to (\ref{eq:SCopSolution}). 
Let us call such a set a {\it complete data-set}. 
For example, in the isotropic case this set can be chosen as follows:
the far-field pattern of the scattered electric field for all angles of observation,
all angles of propagation of the incident time harmonic plane wave, three
linearly independent polarizations and a single fixed frequency \cite{Colton1992a, Colton1992b}.
Another possibly complete data-set may occur when the field due to the original scatterer 
is simulated inside $D$, i.e. $RX=GX$. Although, we are not aware of a rigorous proof of uniqueness
for this rather artificial data-set. The complete boundary measurements all around $D$ may also be 
sufficient \cite{Greenleaf2008}, as well as partial boundary measurements with some 
natural restrictions on the type of contrast functions \cite{Druskin1998, Harrach2009} 
(proofs are available for the static and diffusive regimes only). 
In any case, we may safely conclude that no effective scatterer exists 
on a complete data-set, as only the true scatterer will give the exact data-match.

Thus, an effective scatterer, which matches the fields exactly, can only be found on 
some subsets of a complete data-set, i.e. for partial illuminations and/or observations. 
Due to the sufficient nature of known uniqueness
conditions, however, we cannot generally pinpoint a subset required for an effective scatterer 
to exist.
On the other hand, we can already conclude that, if an effective scatterer
exists for a certain subset of a complete data-set, then either it does not
exist or has a different effective permittivity on the complementary subset.
Indeed, if we could find the same effective scatterer on two or more complementary 
subsets, then the inverse scattering problem would not be unique on the
complete data-set -- a contradiction. This is the reason behind
the dependence of the effective permittivity on the illumination/observation
conditions. It shows that such dependence is a fundamental property
of all effective models, not just the metamaterial slabs.      

\subsection{Non-uniqueness}
Suppose that we were able to find
an exact effective scatterer on some subset of a complete data-set. Is this 
effective scatterer unique? Since we are not sure about the required size of 
the subset, let us consider a very simple situation, where
we have only one incident field and a single data-point.
Then, from (\ref{eq:SingleData}) and (\ref{eq:EffInv}) we obtain
 \begin{align}
 \label{eq:EffInvSingleData}
 \langle X_{\rm ef}[I-GX_{\rm ef}]^{-1}u^{\rm in},r\rangle=
 \langle X[I-GX]^{-1})u^{\rm in},r \rangle=u^{\rm sc}.
 \end{align}
Obviously, any possible non-uniqueness of $X_{\rm ef}$ is a property of the effective 
model itself, not of the particular data-point $u^{\rm sc}$. 
Although such non-uniqueness may be generally anticipated, it is rather
difficult to prove. This is because, unlike the inverse source problem discussed
at the end of the previous section, or inverse scattering in the Born approximation,
the present problem is non-linear. The fact that the operator $R$
has a non-trivial null-space does not yet prove anything.  
Indeed, consider an effective
model with the contrast function of the form $\chi(\bx,\omega)=\chi_{\rm ef}(\omega)\rho(\bx)$, 
where $\rho(\bx)$ is the spatial profile function (in operator notation we shall write $X_{\rm ef}=\chi_{\rm ef}P$).  For a homogeneous
model this profile may be defined as: $\rho(\bx)=1$, $\bx\in D$; $\rho(\bx)=0$, $\bx\notin D$; and $\rho(\bx)$ 
is H{\"o}lder-continuous across the boundary of $D$. 
For this model the problem reduces to finding just one complex
number -- the effective permittivity $\varepsilon_{\rm ef}$ 
or the effective contrast $\chi_{\rm ef}$. If we now apply the Born 
approximation, i.e.,
 \begin{align}
 \label{eq:BornSingle}
 \langle \chi_{\rm ef}P[I-\chi_{\rm ef}GP]^{-1}u^{\rm in},r\rangle\approx 
 \chi_{\rm ef} \langle Pu^{\rm in},r\rangle = u^{\rm sc},
 \end{align}
then the value of the effective contrast is uniquely determined by
a single data-point and a single incident field, and is given by
 \begin{align}
 \label{eq:XefBorn}
 \chi_{\rm ef} = \frac{u^{\rm sc}}{\langle P u^{\rm in},r\rangle}.
 \end{align}
However, if we do not make any approximations, then $\chi_{\rm ef}$ is, in general, 
non-unique. To show this we start by comparing the following eigenvalue problems:
 \begin{align}
 \label{eq:TwoHom1}
 GPu_{\lambda}&=\lambda u_{\lambda},
 \\
 \label{eq:TwoHom2}
 [I-\chi_{\rm ef} GP]u_{\lambda}&=\lambda_{\chi} u_{\lambda}.
 \end{align}
Using the same set of eigenfunctions we deduce the relation
 \begin{align}
 \label{eq:EigGen}
 \lambda_{\chi}=1-\chi_{\rm ef}\lambda.
 \end{align}
Consider incident fields with different spectral content.
First, we assume that our external source creates the field, which
looks like one of the eigenfunctions:
 \begin{align}
 \label{eq:IncOneEig}
 u^{\rm in}=a_{\lambda}u_{\lambda}.
 \end{align}
Now, using (\ref{eq:EffInvSingleData}) and (\ref{eq:TwoHom2})-(\ref{eq:EigGen}), we equate the data from two scatterers, characterized
by $\chi_{\rm ef}$ and $\chi_{\rm ef}'=\alpha \chi_{\rm ef}$, respectively,
 \begin{align}
 \label{eq:IncOneExprAlpha}
 \begin{split}
 \frac{\alpha\chi_{\rm ef}}
 {1-\alpha\chi_{\rm ef}\lambda}
 a_{\lambda}\langle P u_{\lambda},r\rangle = 
  \frac{\chi_{\rm ef}}
 {1-\chi_{\rm ef}\lambda}
 a_{\lambda}\langle Pu_{\lambda},r\rangle.
 \end{split} 
 \end{align}
This leads to the equation
 \begin{align}
 \label{eq:AlphaEq1}
 \begin{split}
 \frac{\alpha}
 {1-\alpha\chi_{\rm ef}\lambda}
 = 
 \frac{1}
 {1-\chi_{\rm ef}\lambda},
 \end{split} 
 \end{align}
which has only one solution, $\alpha=1$. 
Thus, with a single-mode incident field this effective model
is unique, i.e. $\chi_{\rm ef}$ is uniquely determined by a single
data-point $u^{\rm sc}$.
However, a realistic incident field, e.g. a plane wave or the field of a dipole source,
will contain many different modes $u_{\lambda}$. Consider, for example, the incident field of the form
 \begin{align}
 \label{eq:IncTwoEig}
 u^{\rm in}=a_{1}u_{1}+a_{2}u_{2},
 \end{align}
where both $u_{1}$ and $u_{2}$ are eigenfunctions. Then, we obtain
the following problem for $\alpha$:
 \begin{align}
 \label{eq:IncTwoExpr1}
 \begin{split}
 \frac{\alpha c_{1}}
 {1-\alpha\lambda_{1}'}
 +
 \frac{\alpha c_{2}}
 {1-\alpha\lambda_{2}'}
 =\frac{c_{1}}
 {1-\lambda_{1}'}
 +
 \frac{c_{2}}
 {1-\lambda_{2}'},
 \end{split} 
 \end{align}
where $c_{1,2}=a_{1,2}\langle Pu_{1,2},r\rangle$ and $\lambda_{1,2}'=\chi_{\rm ef}\lambda_{1,2}$.
This problem reduces to a quadratic equation with two roots:
 \begin{align}
 \label{eq:TwoAlphasSolution}
 \begin{split}
 \alpha_{1}&=1,
 \\
 \alpha_{2}&=\frac{c_{1}+c_{2}-c_{1}\lambda_{2}'-c_{2}\lambda_{1}'}
 {c_{1}\lambda_{2}'+c_{2}\lambda_{1}'-c_{1}(\lambda_{2}')^{2}-c_{2}(\lambda_{1}')^{2}},
 \end{split}
 \end{align}
showing that the effective model is non-unique. 
The location of the second solution depends on the exact balance of the modes in the incident field,
location of the receiver, and the eigenvalues, 
which in their turn depend on the applied frequency and the geometry of the object.
In general, there will be as many solutions as there are modes present in the incident field. 
And we may expect them all to be different, if the modes correspond to distinct eigenvalues.

\subsection{Approximate effective permittivity}
Above we have considered data from a homogeneous object and showed that the 
homogeneous model itself cannot always be uniquely inverted. One can expect, however,
that addition of just a few incident fields and/or measurement points will cure that problem.
Hence, it would seem a reasonable strategy to add, say, new receivers when we
deal with the data from an inhomogeneous object as well, thus obtaining a unique
value of the effective permittivity. However, here we run into the problem of non-existence
discussed above. Although, an effective permittivity with a single incident field
and a single data-point always exists (this can be proven using the technique of 
the previous subsection), it is different for different observation points
and incident fields (recall our discussion on complementary data-sets). 
Hence, in general, no single effective permittivity will fit all the data, even if it were just 
two data-points and a single incident field.

This brings us back to the original purpose of the effective permittivity
as a simplifying approximation. While the exact effective permittivity appears
not to exist, we can still find an approximate effective permittivity which
minimizes the discrepancy between the scattered fields. We have to realize, however, that this discrepancy will generally grow
as we add new data to the problem. From being exactly zero for many different values of
$\varepsilon_{\rm ef}$ to some finite value at some possibly unique
minimum, which corresponds to the approximate effective permittivity.
Whether that minimum is definitely unique -- we do not know.

Further, since the location of exact effective permittivities, obtained
for each separate data-point, depends
on the eigenvalues of the scattering operator, which in their turn depend
on the geometry of the object, we expect the value(s) of an approximate 
effective permittivity to be geometry-dependent as well.

\subsection{Singularities}
A popular way of enforcing uniqueness on the effective permittivity is by choosing a ``physical'' value
of $\varepsilon_{\rm ef}$ as opposed to other, ``non-physical'' ones. However, when constructing 
an effective model all we should care about is the match in the scattered fields and the 
well-posedness of the effective model. From this point of view, some currently disregarded
values of $\varepsilon_{\rm ef}$  are perfectly acceptable. For example, effective permittivities
with the negative imaginary part are often dismissed as representing a ``non-physical'' medium 
with gain. First of all, pumped media with population inversion are no less physical than ordinary
lossy materials, being, in fact, less exotic then the negative refraction media. Thus, the effective
model with gain, if it matches the scattered field data, should not be dismissed simply because 
one does not like it. There are certain values of the effective permittivity, though, which make the
effective model ill-posed and should be disregarded. These are the points of perfect resonances
where the spectrum of the effective scattering operator $I-GX_{\rm ef}$ contains $\lambda=0$.
In \cite{Budko2006b} it was shown that a perfect eigenvalue-based resonance does occur in media with gain.
However, this happens only for some specific combinations of the permittivity, frequency, and geometry.
This is what one strives for in the design of laser cavities. Hence, only certain ``discrete'' values
of $\varepsilon_{\rm ef}$ with ${\rm Im}\,\varepsilon_{\rm ef}<0$ will cause the trouble.

A question of specific importance to metamaterials is whether real negative values of $\varepsilon_{\rm ef}$
are acceptable. In \cite{Budko2006b} it was argued that in this case an essential-spectrum-based perfect 
resonance may occur rendering the effective model ill-posed. 
Indeed, the essential spectrum is $\lambda_{\rm es}=\varepsilon(\bx,\omega)/\varepsilon_{0}$, 
$\bx\in{\mathbb R}^{3}$, and in the H{\"o}lder-continuous case one needs to connect somehow the real unit 
with a negative real value. Thus, if we use the shortest path, it will pass trough the zero of the complex plane. 
However, in principle, it is possible to create or imagine a scatterer whose permittivity 
avoids the $\varepsilon(\bx,\omega)=0$ point on its way from the real unit to the chosen negative real value. 
For instance, an object with a thin lossy outer coating seems to be acceptable from that point of view. 
There is another problem however, which was not anticipated in \cite{Budko2006b}.
It turns out that there is also a potentially infinite number of discrete real values of $\varepsilon_{\rm ef}$
spread from zero to negative infinity, which all cause perfect eigenvalue-based resonances and make
the effective model ill-posed. Indeed, consider an effective homogeneous scatter with the
contrast $\chi_{\rm ef}(\bx,\omega)=P$, i.e. $\varepsilon_{\rm ef}/\varepsilon_{0}=2$. Then,
 \begin{align}
 \label{eq:EigOne}
 \left\Vert [I-GP-\lambda I]^{-1}\right\Vert=\frac{1}{\vert 1-\lambda\vert}\Vert [I-\frac{1}{1-\lambda}GP]^{-1}\Vert=\infty,
 \end{align}
if $\lambda$ is an eigenvalue of $I-GP$. Therefore, for a general homogeneous effective scatterer
with permittivity $\varepsilon_{\rm ef}$ we have
 \begin{align}
 \label{eq:EigEps}
 \left\Vert [I-GX_{\rm ef}]^{-1}\right\Vert=\left\Vert [I-(\varepsilon_{\rm ef}/\varepsilon_{0}-1)GP]^{-1}\right\Vert=\infty,
 \end{align}
if
 \begin{align}
 \label{eq:EpsEig}
 \frac{\varepsilon_{\rm ef}}{\varepsilon_{0}}=1+\frac{1}{1-\lambda},
 \end{align}
where $\lambda$'s depend only on the geometry of $D$ and the applied frequency.
Thus, no matter what kind of illumination and/or scattering data we use in
(\ref{eq:EffInv}), the effective homogeneous 
model will be ill-posed for the above values of effective permittivity.

Let us figure out where these values are on the complex plane. To do so we
need to know the location of $\lambda$'s for a homogeneous object with permittivity
$\varepsilon_{\rm ef}/\varepsilon_{0}=2$. They will generally occupy the lower
half of the complex plane, i.e. ${\rm Im}\,\lambda \le 0$. Then, from (\ref{eq:EpsEig})
we conclude that the troublesome values of $\varepsilon_{\rm ef}/\varepsilon_{0}$
will also be in the lower half of the complex plane. Thus we confirm the anticipated problems 
for some discrete values of $\varepsilon_{\rm ef}$ exhibiting ``gain'', i.e. 
${\rm Im}\,\varepsilon_{\rm ef} < 0$. There are, however, also purely real $\lambda$'s.
For example, in \cite{Budko2005}, it was shown that in the quasi-static case 
$\omega\rightarrow 0$, i.e., for objects much smaller than the wavelength of the incident
field, all discrete eigenvalues will be concentrated inside the convex hull of the 
essential spectrum. Since the essential spectrum is now a segment of real line
stretching between one and two (provided that the H{\"o}lder-continuous $\rho(\bx)$
is chosen correspondingly), that is where all discrete eigenvalues will be distributed
in the quasi-static regime. From (\ref{eq:EpsEig}) we conclude that in that  case
there will be infinitely many discrete real values of $\varepsilon_{\rm ef}/\varepsilon_{0}$
spread all over the interval $(-\infty,0]$, such that the effective model is ill-posed.
We expect (see numerical examples in \cite{Budko2006a, Budko2006b} and here) 
that for our base object with $\varepsilon_{\rm ef}/\varepsilon_{0}=2$ 
real eigenvalues are present at higher frequencies as well,
and that the above conclusion is not specific to small objects.

\subsection{Effective models of higher complexity}
In a recent review article \cite{Greenleaf2008} an explicit connection has been made between the
inverse scattering (Calder{\'o}n) problem of reconstructing an object from boundary measurements
and the problem of invisibility, discussed in the context of metamaterials. 
Metamaterials are believed to produce almost arbitrary anisotropic functions of effective permittivity and
permeability. Since another well-known result in inverse scattering theory
tells us that anisotropic objects cannot be uniquely reconstructed from boundary measurements \cite{Greenleaf2008},
the question arises about the possibility of employing effective models that in some aspects 
are more complex than the original scatterer. For example, a composite object
consisting of isotropic parts could be modelled as an anisotropic one.

Thus, in terms of the effective inversion problem (\ref{eq:EffInv}) we are now interested in the
possibility of $X_{\rm ef}$ being a more complicated operator than the original $X$,
so that a data-set which is complete for the original scatterer is incomplete for the effective 
model. If only the permittivity is considered to be anisotropic, then $X_{\rm ef}$ becomes a three-diagonal
multiplication operator. If a magnetic contrast is present as well, then the forward scattering 
problem will look like
 \begin{align}
 \label{eq:ElectricMagnetic}
 \left(
 \begin{bmatrix}
 I& 0\\
 0 & I
 \end{bmatrix}
 - 
 \begin{bmatrix}
 G_{11} & G_{12}\\
 G_{21} & G_{22}
 \end{bmatrix}
  \begin{bmatrix}
 X_{\rm e} & 0\\
 0 & X_{\rm m}
 \end{bmatrix}
 \right) 
 \begin{bmatrix}
 e\\
 h
 \end{bmatrix}
 =
 \begin{bmatrix}
 e^{\rm in}\\
 h^{\rm in}
 \end{bmatrix},
 \end{align}
where $G_{kl}$ are known integral operators \cite{Samokhin2001}, 
$X_{\rm e, m}$ are three-diagonal contrast-operators, and $e$ and $h$ are the
electric and magnetic fields, respectively. Uniqueness of the solution is 
guaranteed, if at least one of the constitutive parameters has losses or,  in the lossless
case, all components 
of all tensors are three times continuously differentiable functions of coordinates.
The necessary and sufficient condition on the existence of the solution of (\ref{eq:ElectricMagnetic})
is also derived in \cite{Samokhin2001} for H{\"o}lder-continuous tensor-valued functions
of permittivity and permeability, and requires
 \begin{align}
 \label{eq:ExistenceAnisotropic}
 \begin{split}
 \sum\limits_{k=1}^{3}\sum\limits_{m=1}^{3}\theta_{k}\varepsilon_{km}(\bx,\omega)\theta_{m}\ne 0,
 \\
 \sum\limits_{k=1}^{3}\sum\limits_{m=1}^{3}\theta_{k}\mu_{km}(\bx,\omega)\theta_{m}\ne 0,
 \\
 \bx\in{\mathbb R}^{3},
 \end{split}
 \end{align}
where $\theta_{k}$, $k=1,2,3$ are the Cartesian components of an arbitrary real vector of length one.
Obviously, this condition will cause singularities in the inverse scattering problem not just along
lines and curves as in the isotropic case, but over whole areas of the complex plane. In particular, 
equation (\ref{eq:ExistenceAnisotropic}) seems to exclude the values of permittivity and permeability 
required for perfect invisibility \cite{Greenleaf2008}.

When considering effective models of higher complexity we do not have to limit ourselves 
to anisotropy. Why not consider spatial dispersion or even extra spatial dimensions?
While these models may seem over the top, they simply illustrate the main problem
with this approach. We agree with the authors of \cite{Menzel2008a, Menzel2008b, Menzel2009} that, in general,
effective models of higher complexity contradict the basic goal of effective modelling, 
which was to simplify the original scattering model.
We also doubt that these models can give us any useful information about various exotic phenomena
as suggested, for example, in \cite{Greenleaf2007}. Indeed, if a higher complexity effective model exists, then
it must be, by definition, completely equivalent to the original composite, but mundane scatterer.
Conversely, any electromagnetic effect which makes this exotic model identifiable, i.e., different
from the original scatterer, will violate the existence condition on the effective model.

In fact, it is quite easy to come up with a higher complexity model,
which seemingly shows physics, not present in the original object. 
For instance, we know that the value of effective permittivity for a homogeneous model generally varies 
with the incident field. Hence, we can say that $\varepsilon_{\rm ef}$, which 
solves (\ref{eq:EffInv}), depends on $u^{\rm in}$. Although, multiplying the incident field by 
a constant does not change the value of $\varepsilon_{\rm ef}$, adding another source to 
the existing one will cause a variation in $\varepsilon_{\rm ef}$. Mathematically, this
can be expressed as
 \begin{align}
 \label{eq:Nonlinearity}
 \begin{split}
 \varepsilon_{\rm ef}[Au^{\rm in}_{1}]&=\varepsilon_{\rm ef}[u^{\rm in}_{1}],
 \\
 \varepsilon_{\rm ef}[u^{\rm in}_{1}+u^{\rm in}_{2}]&
 \ne \varepsilon_{\rm ef}[u^{\rm in}_{1}],
 \end{split} 
 \end{align}
showing that $\varepsilon_{\rm ef}$ is a {\it nonlinear} functional of the incident field. 
Thus a linearly reacting inhomogeneous object may be viewed, to a certain extent, as a 
nonlinear homogeneous scatterer.

Finally, the present isotropic single-frequency model, where the effective permittivity 
is allowed to vary with frequency, even if the permittivity of the original scatterer does not, 
is, obviously, also a model of higher complexity, if considered over the range of frequencies.
Whether addition of multiple frequencies to the partial illumination/observation data-set
makes it a complete data-set is an open question. It probably does, since a considerable 
improvement in the reconstruction of spatially inhomogeneous Lorentz-type dispersive media 
was observed in \cite{Budko2002} with the addition of just a few frequencies.

\section{Finding effective permittivity}
It is well-known that an analytical solution for an effective 3D scatterer of general 
geometry $D$ and/or general profile $\rho(\bx)$ is not available. 
Numerical solution of the effective inversion problem (\ref{eq:EffInv})
requires an inverse of a $N \times N$ matrix obtained after discretization of the integral
operator featured in equation (\ref{eq:IntegralEquation}). This is practically impossible
for an electrically large object, as $N$ is in the order of millions. 
Yet, the numerical solution of the forward scattering 
problem (\ref{eq:IEop}) for some particular $u^{\rm in}$ can be found with 
an iterative method. In the effective inversion problem, however,
such solutions must be found for many values of the effective permittivity.
Thus, if solution takes $M$ iterations and we need to find it for $K$
values of $\varepsilon_{\rm ef}$, then we would need to carry out
$MK$ iterations. Each iteration is computationally roughly equivalent 
to a matrix-vector product. Although the special symmetries of the 
integral operator $G$ allow for a significant simplification of this process
via the FFT algorithm, we still have a serious computational bottleneck here.

In principle, we could try to minimize the number of trial $\varepsilon_{\rm ef}$'s
when searching for the one which minimizes the data-discrepancy.
For example, we could apply a Newton-type minimization algorithm.
However, this algorithm is good for finding the minimum of a single-minimum 
functional, and should not be used with multi-minima problems,
as the one at hand here. Not to mention that we would certainly prefer to visualize 
the norm of the data-discrepancy for $\varepsilon_{\rm ef}$ in a certain range 
and confirm the analytical predictions made in this paper. 

An algorithm, which circumvents the computational bottleneck of the
effective inversion problem, was proposed in \cite{Budko2004}.
It exploits the invariance of the Krylov subspace constructed
by the Arnoldi algorithm with respect to a shift by the parameter 
$\chi_{\rm ef}$. Practically this means that the computationally 
expensive iterations will be carried out only once, say for $\chi_{\rm ef}=1$,
and the resulting Arnoldi-vectors and the associated $M\times M$ upper Hessenberg matrix 
may be re-used with different values of $\chi_{\rm ef}$. Thus, instead of carrying 
out $MK$ iterations with the problem of the order $N\times N$, we
have to perform $M$ iterations with the problem of the order $N\times N$ 
and find solutions of $K$ linear problems of the order $M\times M$. 
Since typically $M \ll N$, the total number of operations is now much smaller,
and we have what is called a {\it reduced-order} algorithm.

In this section we shall present numerical examples for the following three objects:
a homogeneous object, an elementary cell of a dielectric photonic crystal, and
a larger rectangular piece of this crystal. In all cases we shall visualize
the discrepancy between the measured (here -- simulated) scattered field from a true
object and the field from a homogeneous effective model of the same outer shape.
Namely, we shall plot the $\log_{10}$ of the following functional:
 \begin{align}
 \label{eq:Functional}
 \begin{split}
 &F\left[\frac{\varepsilon}{\varepsilon_{0}}\right]=
 \\
 &\frac{\Vert u^{\rm sc} - 
 \left(\frac{\varepsilon}{\varepsilon_{0}}-1\right)
 RP\left[I-\left(\frac{\varepsilon}{\varepsilon_{0}}-1\right)GP\right]^{-1}u^{\rm in}
 \Vert^{2}}{\Vert u^{\rm sc}\Vert^{2}},
 \end{split} 
 \end{align}
representing a two dimensional non-negative function of the real and 
imaginary parts of $\varepsilon/\varepsilon_{0}$. In all our examples
we focus on a single data-point case, i.e., $u^{\rm sc}$ is just a complex number (both the amplitude and
the phase of the field are presumed to be measured). Hence, the norm in (\ref{eq:Functional})
is simply the square of the absolute value of the discrepancy.
If one wishes to know what would happen with more data-points included in the calculations, 
where the norm becomes a sum of the single data-point norms, then
one simply has to add the corresponding functionals (\ref{eq:Functional}).

\begin{figure*}[t]
\hspace*{0cm}\epsfig{file=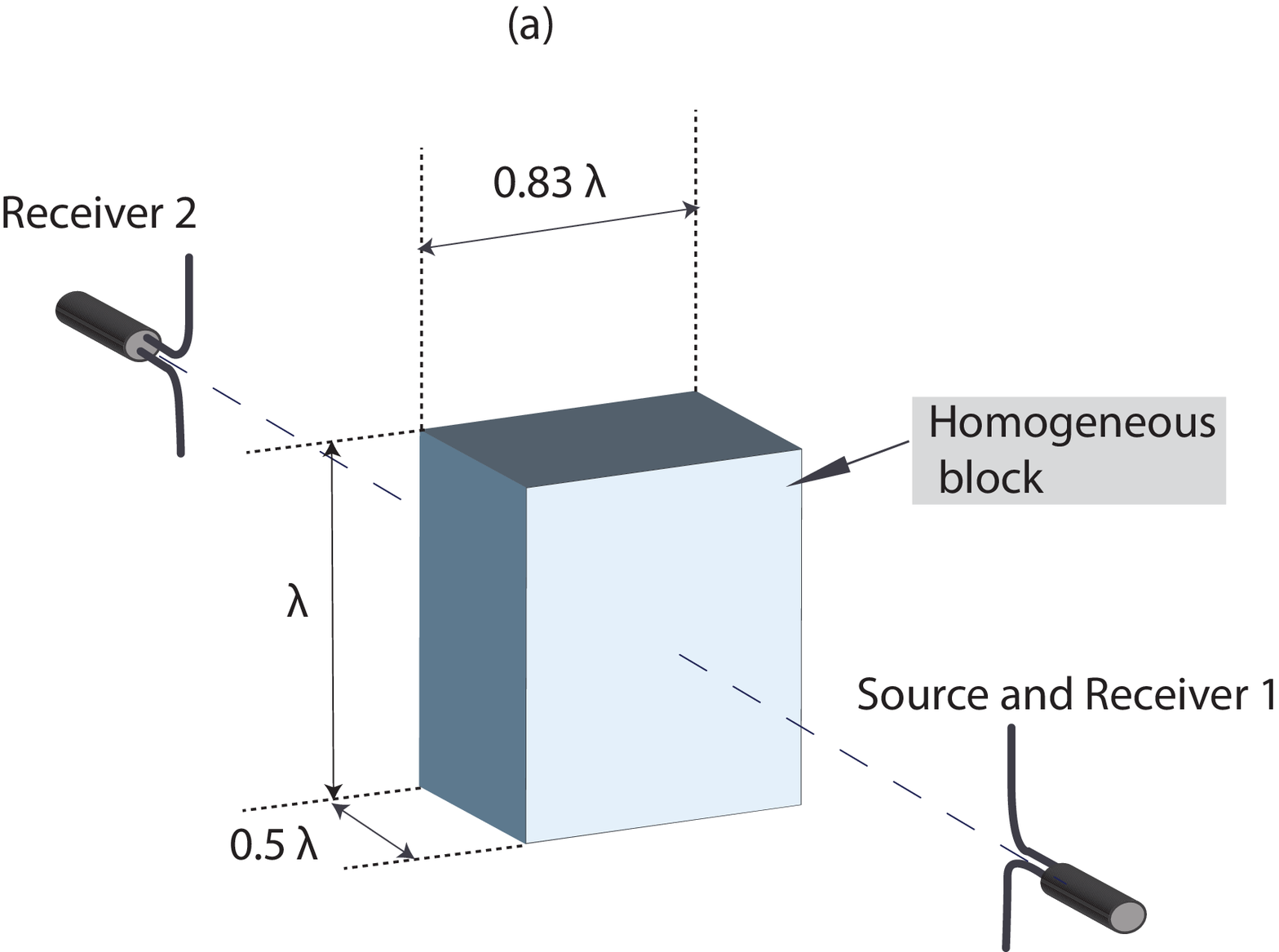,width=7cm}
\hspace*{2.5cm}\epsfig{file=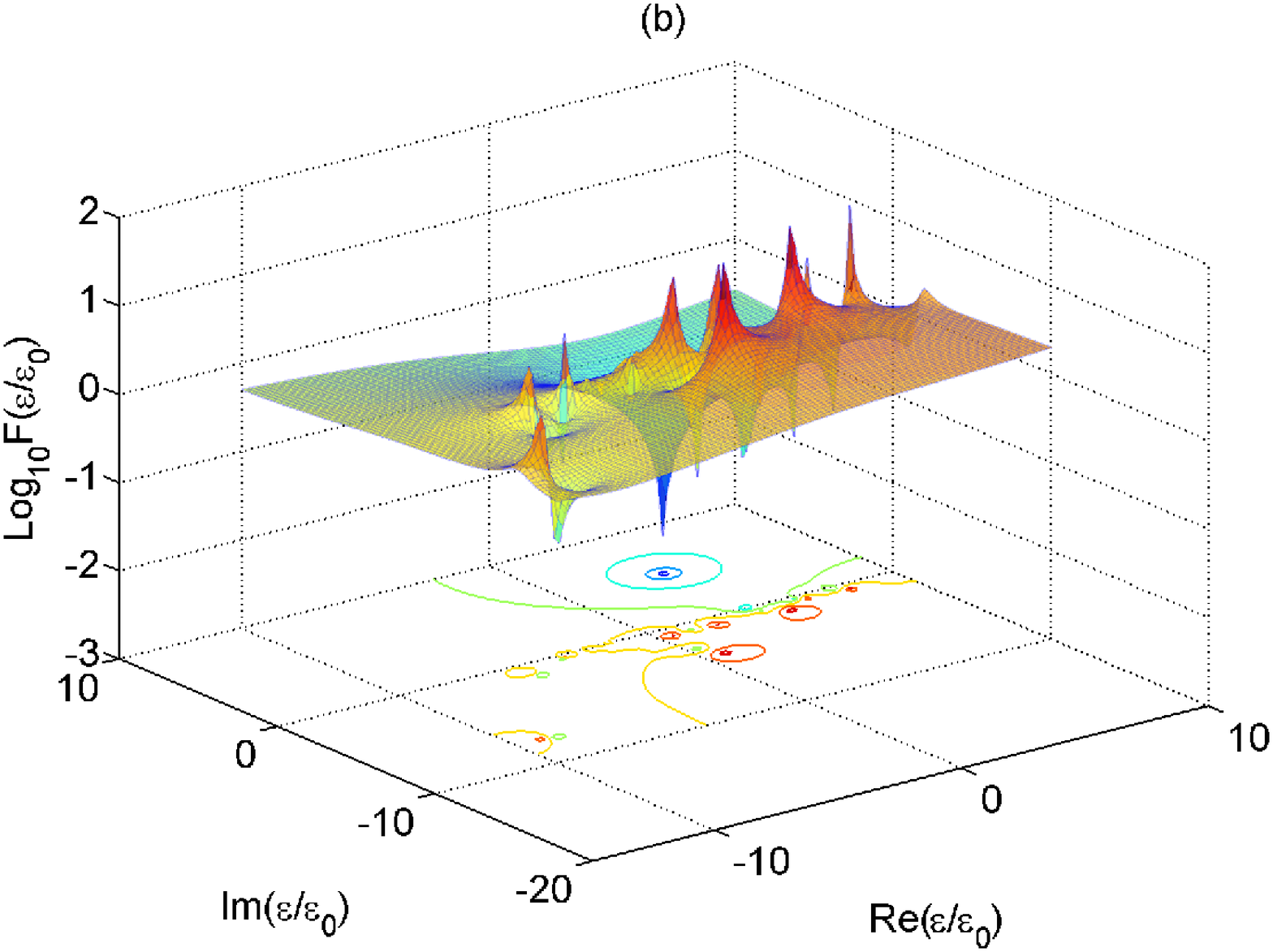,width=7cm}
\\ \vspace*{0.5cm}
\hspace*{0cm}\epsfig{file=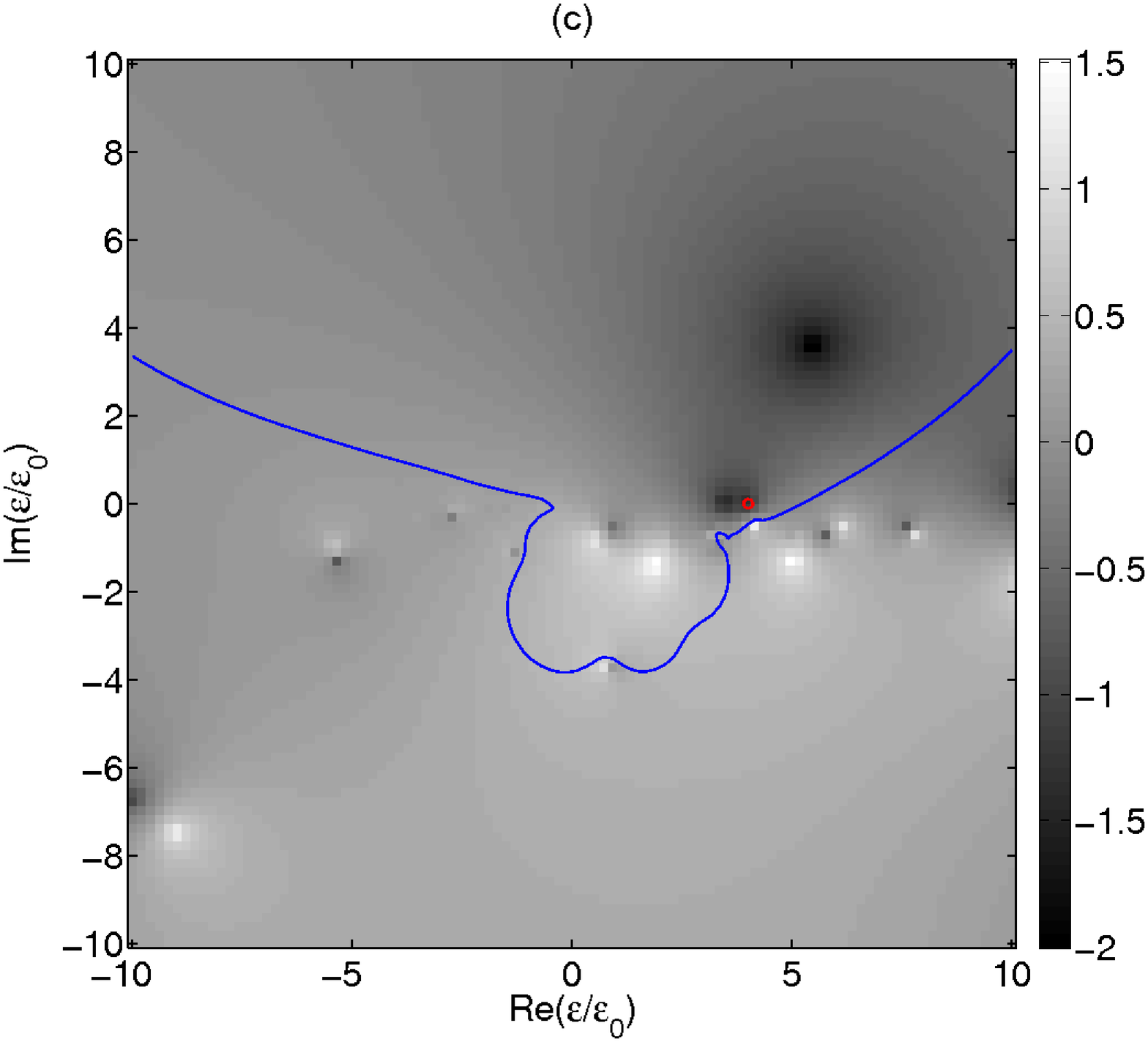,width=7cm}
\hspace*{2.5 cm}\epsfig{file=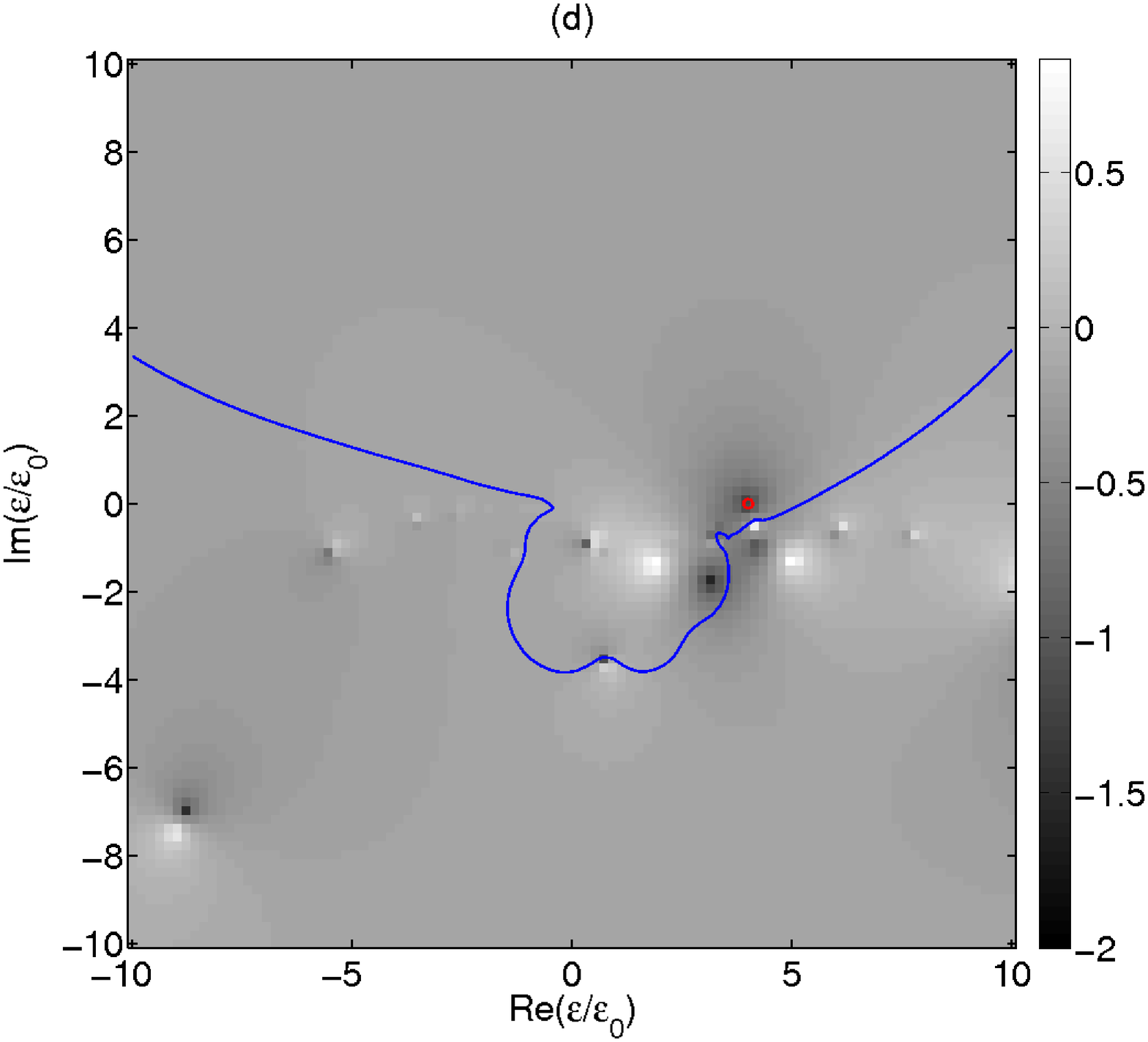,width=7cm}
\caption{Finding the effective permittivity of a homogeneous object:
(a) -- scattering configuration; (b) -- surface of the data-discrepancy functional for Receiver~1;
(c), (d) -- data-discrepancy functionals as two-dimensional images for Receivers~1 and 2. 
Non-uniqueness of the homogeneous model is clearly visible in (c). 
The presence of a ``stationary'' minimum (circle) means existence of the exact effective
permittivity, which is simply the true permittivity of the homogeneous object $\varepsilon/\varepsilon_{0}=4$.}
\end{figure*}

First we consider a homogeneous rectangular object of a resonant size depicted in Fig.~1~(a).
The dimensions in terms of the free-space wavelength are indicated in the figure. The source of the incident field is an electric 
point-dipole situated one wavelength away from one the object faces. The dipole is polarized as shown in the figure.
We simulate a single Cartesian component (indicated by the orientation of the receivers in the figure) 
of the scattered field, one wavelength away from the two opposite sides of the object, simulating the transmission and 
the backscattering measurement setups. 
The relative permittivity of the true object is $\varepsilon/\varepsilon_{0}=4$. 
The effective scatterer is also homogeneous and has the
same outer shape. The discrepancy between the scattered field of the true object and the field scattered by
homogeneous objects with other values of relative permittivity is shown as a surface in Fig.~1~(b),
where the horizontal axes are the real and the imaginary parts of the trial relative permittivities,
and the height of the surface gives the value of the discrepancy functional for Receiver~1. To visualize this 
functional on a $100\times 100$ grid we have computed the data-discrepancy for $K=10000$ different values of 
the complex relative permittivity, which would not be possible without a reduced-order algorithm discribed above.

As expected from our theoretical analysis, the functional has many maxima and minima. 
Minima correspond to the match between the data, hence, giving the effective permittivities. 
Maxima, correspond to the singularities of (\ref{eq:Functional}), which, as predicted by equation (\ref{eq:EpsEig}),
are situated in the lower half of the complex $(\varepsilon/\varepsilon_{0})$-plane and along the negative real axis.
It is easier to analyze the shape of the functional on a two-dimensional image as the ones
shown in Fig.~1~(c), (d), where the height of the functional is now represented by the brightness of the pixel.
In these images bright spots correspond to singularities, and dark spots are the effective permittivities. 

Also shown is a contour around an area of the $(\varepsilon/\varepsilon_{0})$-plane where
the forward scattering problem is solved with sufficient accuracy. The norm of the residual of the forward
problem is smaller than 0.01 inside the contour (here, above the curve), 
i.e. we have beteer than 1\% accuracy there.
The variation of accuracy with permittivity is the result of a fixed number 
of Arnoldi iterations applied everywhere (here and below we use $M=60$ iterations). 
We achieve an almost machine precision 
for small contrasts, i.e. for $\varepsilon/\varepsilon_{0}$ in the neighborhood of one,
but arrive at progressively larger residuals for larger contrasts. Also, we cannot count on any good
accuracy in the immediate neighborhood of discrete singularities and for all non-positive real $\varepsilon/\varepsilon_{0}$,
as the forward problem becomes numerically ill-conditioned there. Achieving acceptable accuracy 
for larger areas of the $(\varepsilon/\varepsilon_{0})$-plane comes at a cost of more Arnoldi
iterations, and in our case is limited to $M=60$ by the available computing resources.
On the other hand, the depicted contour gives the norm of the residual of the forward problem,
i.e. the total error in the solution all over the spatial doain $D$ and for all Cartesian components
of the electric field. Whereas, what we should be concerned about is the error in the 
computed scattered field at the receiver location only. That error is probably much smaller
than the norm of the residual on $D$. Therefore, we may expect that the computed reduced-order functionals 
can be trusted way beyond the outlined contour.

The different minima of the discrepancy functional, some in the ``physical'' part of the $(\varepsilon/\varepsilon_{0})$-plane,
see Fig~1~(c), illustrate the non-uniqueness of the homogeneous effective model.
If we compare Fig.~1~(c) and Fig.~1~(d), which correspond to different locations of the receiver, then we notice
that all singularities remain at the same points, since they depend only on the
shape of the scatterer and the applied frequency. At the same time all but one minima
have changed their locations. The one which did not move was, of course, the permittivity
of the original homogeneous object $\varepsilon/\varepsilon_{0}=4$. This gives us a
way to retrieve the unique permittivity of a homogeneous scatterer from just two data-points
via detection of a stationary minimum.

Application of the homogeneous effective model to an inhomogeneous elementary cell,
similar to those used in photonic crystals \cite{Krowne2007}, is depicted in Fig.~2.
We have two cylindrical rods, both with the relative permittivity $\varepsilon/\varepsilon_{0}=4$,
and a $\lambda/3$ gap between their centers, see Fig.~2~(a). 
\begin{figure*}[t]
\hspace*{0cm}\epsfig{file=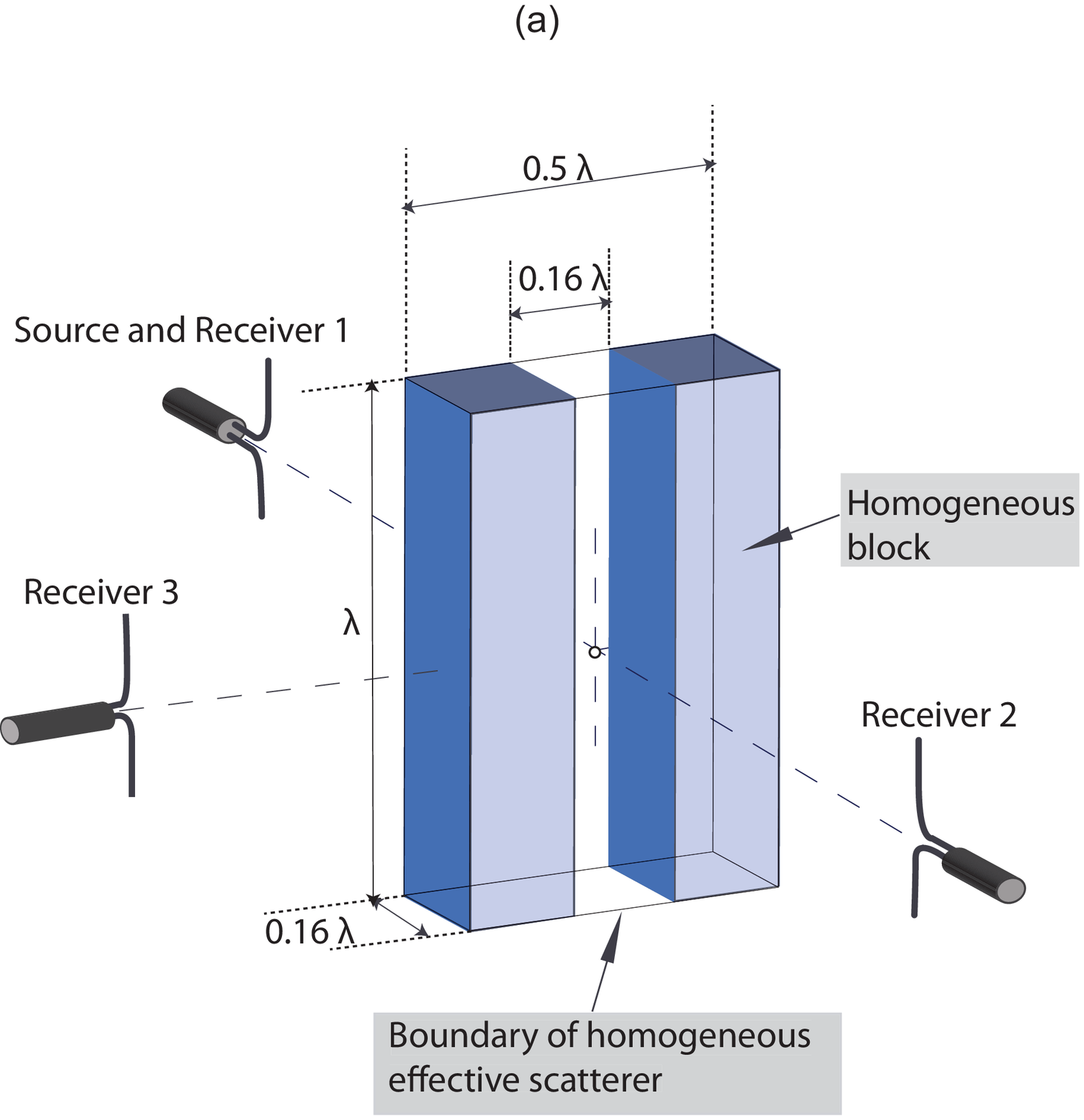,width=7cm}
\hspace*{2.5cm}\epsfig{file=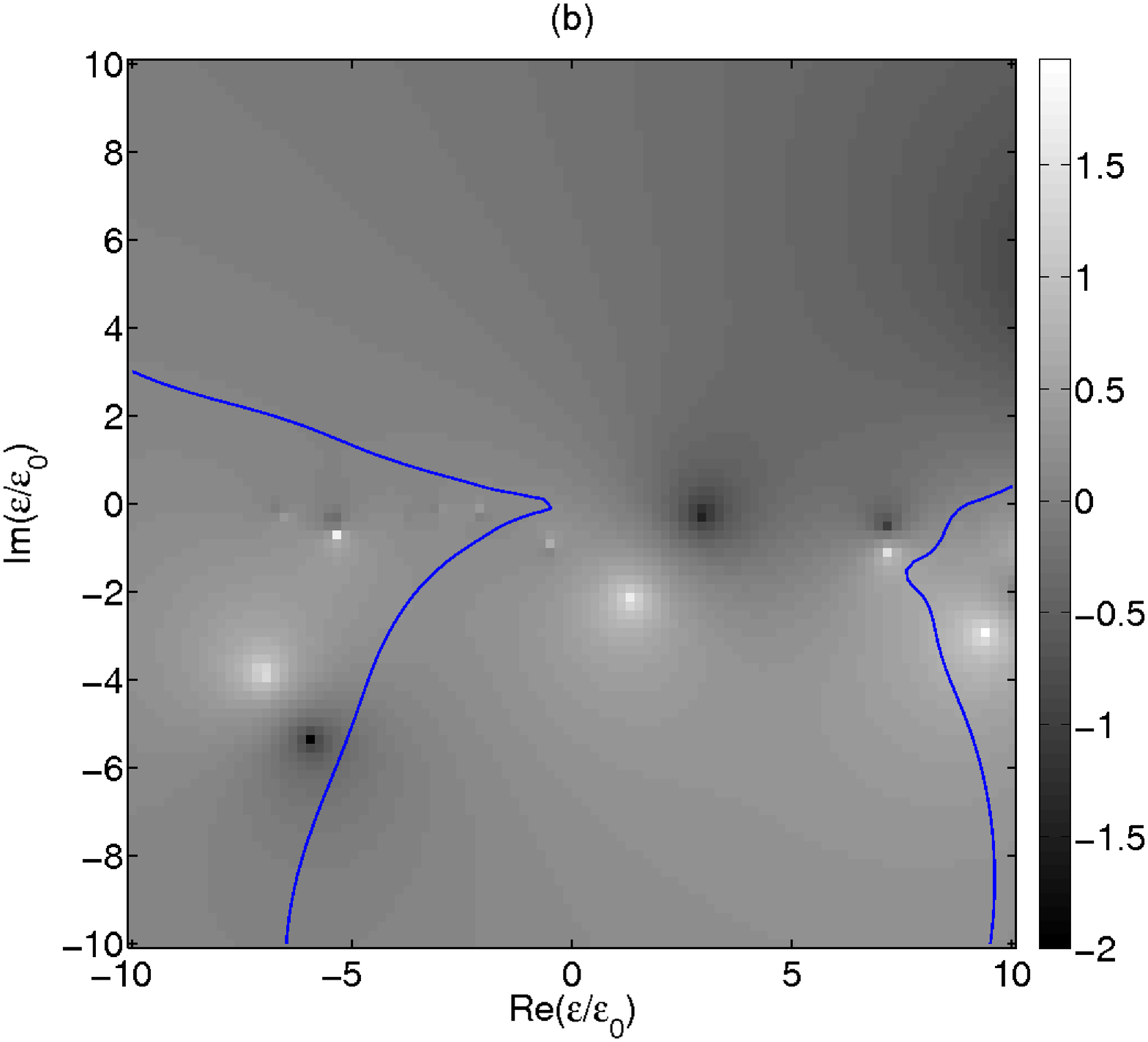,width=7cm}
\\ \vspace*{0.5cm}
\hspace*{0cm}\epsfig{file=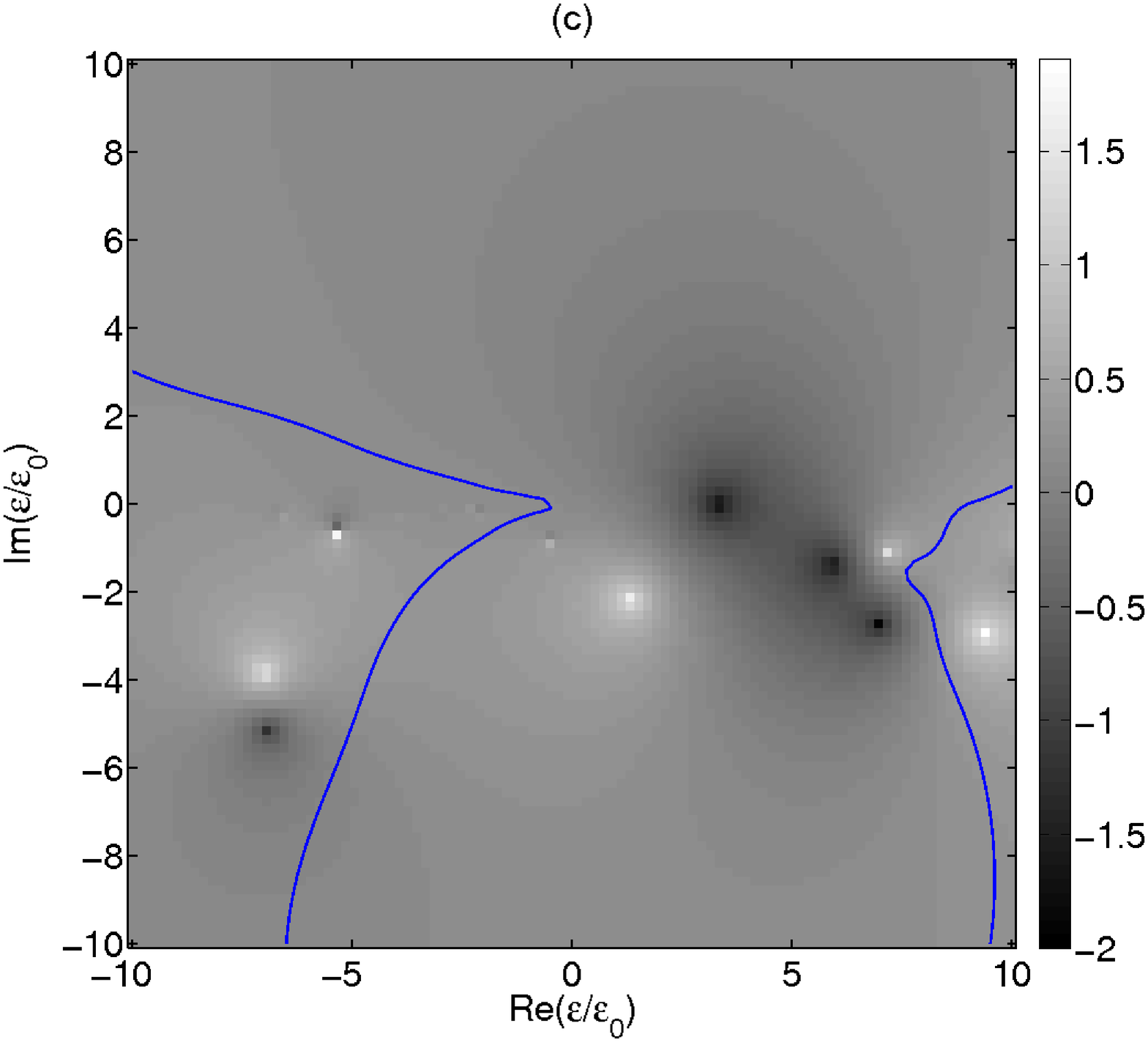,width=7cm}
\hspace*{2.5 cm}\epsfig{file=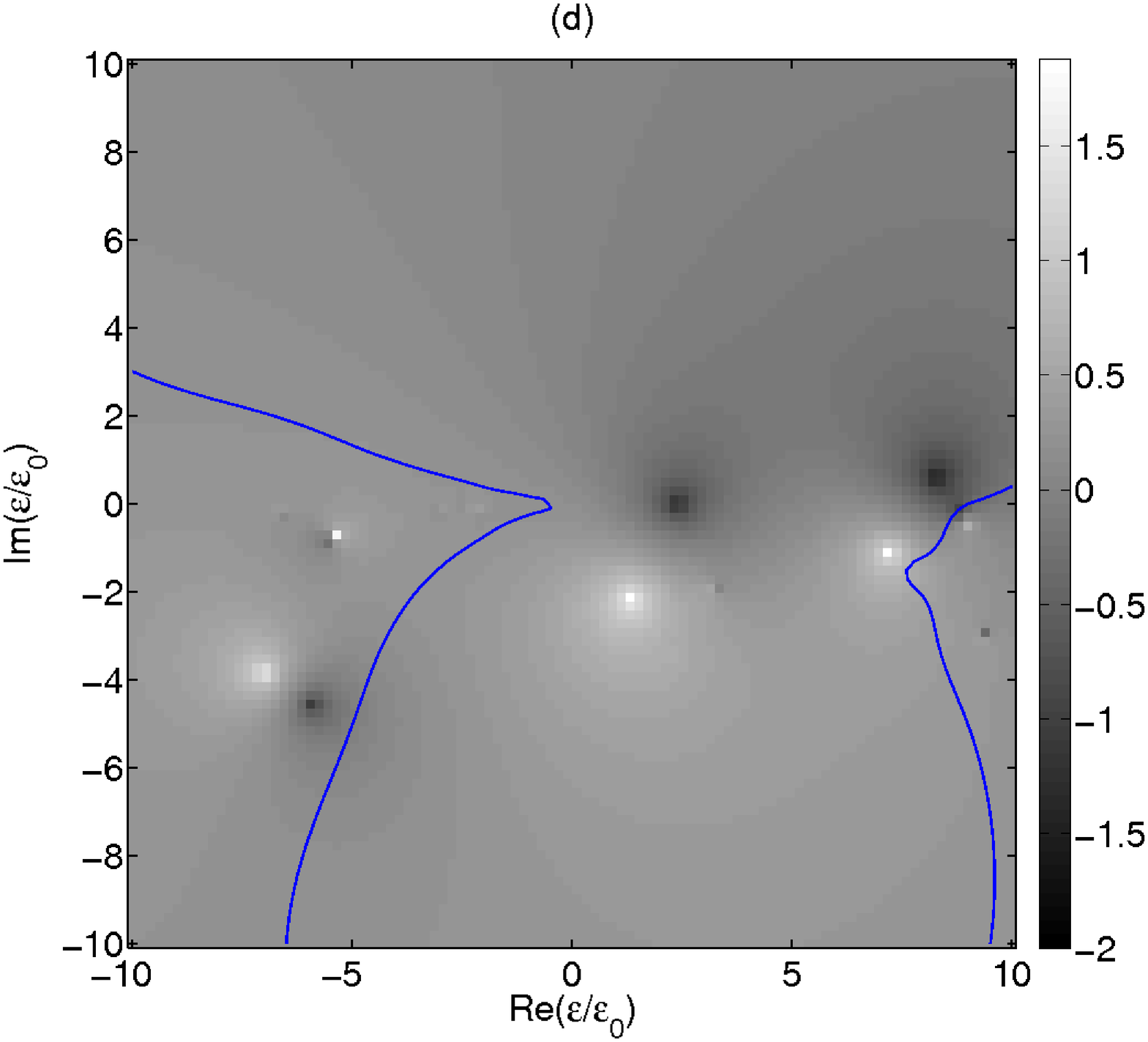,width=7cm}
\caption{An attempt to find the effective permittivity of an inhomogeneous object:
(a) -- scattering configuration; (b), (c), and (d) -- data-discrepancy functionals for Receiver~1,
Receiver~2, and Receiver~3, correspondingly. Notice the presence of two ``physical'' solutions in (b) and (d), 
demonstrating the non-uniqueness, and the absence of a ``stationary'' minimum in (b)--(d), demonstrating
the non-existence of the exact effective permittivity for inhomogeneous objects.}
\end{figure*}
We consider three different locations of the receiver, indicated in Fig.~2-a as
Receiver~1, Receiver~2, and Receiver~3, correspondingly. The distance of all receivers
to the nearest faces of the scatterer is set to one wavelength.  Both the  
true and the effective objects are smaller with respect to the wavelength 
than in the previous example (in the horizontal cross-section).
Hence, we see less singularities and minima in the images of Fig.~2~(b), (c), and (d).
The other significant difference is the absence of a minimum, stationary
with respect to the changes in the measurement setup.  
This illustrates the anticipated non-existence of the
exact effective permittivity in the inhomogeneous case. 

Finally we consider a larger sample of a dielectric photonic crystal, similar to the one
investigated in \cite{Krowne2007} in relation to the phenomenon of negative refraction.
In Fig.~3~(a) the scattering configuration is shown, where all three receivers are one
wavelength away from the corresponding faces of the rectangular effective object.
This effective model has, in fact, the same dimensions as the one considered in Fig~1. 
\begin{figure*}[t]
\hspace*{0cm}\epsfig{file=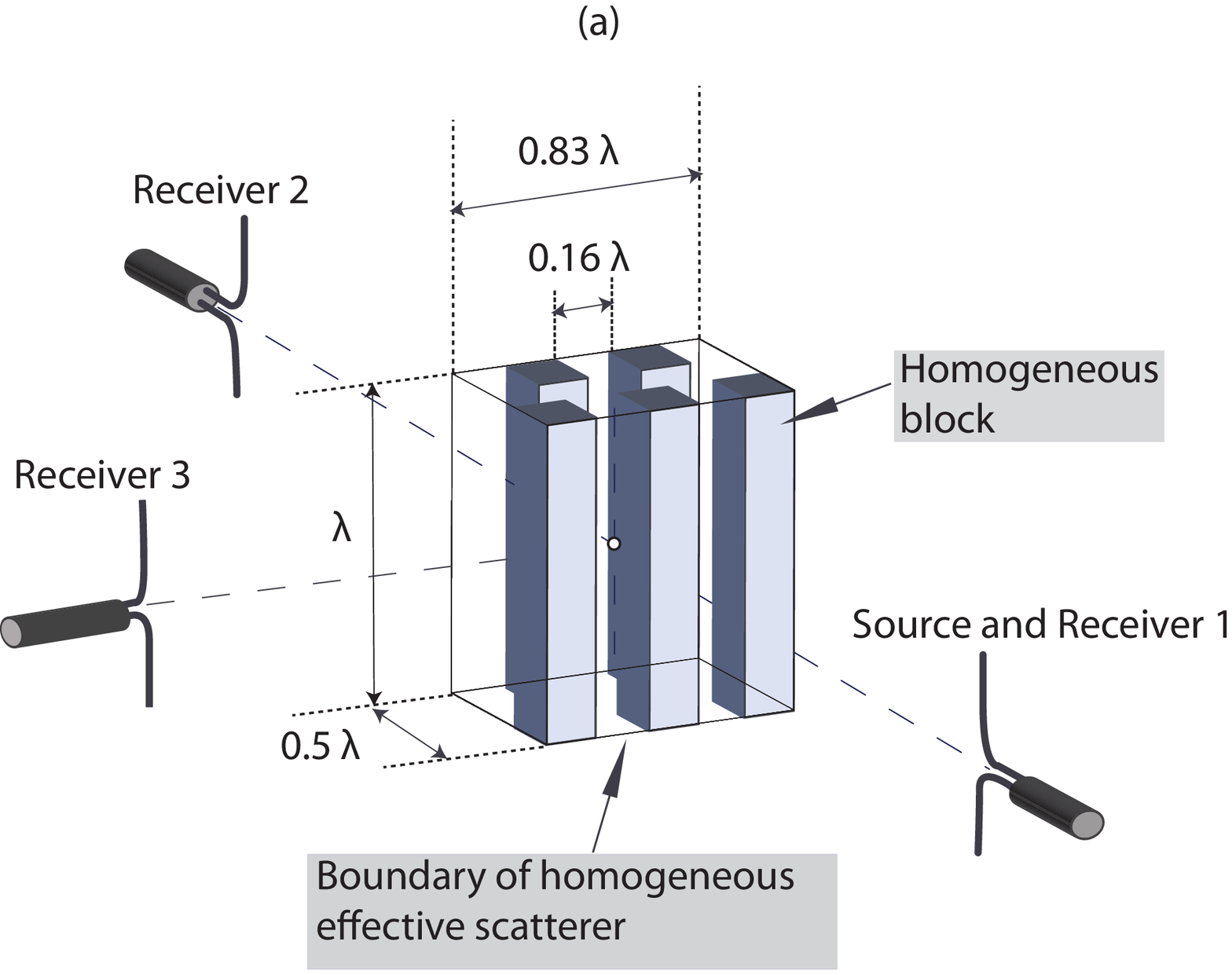,width=7cm}
\hspace*{2.5cm}\epsfig{file=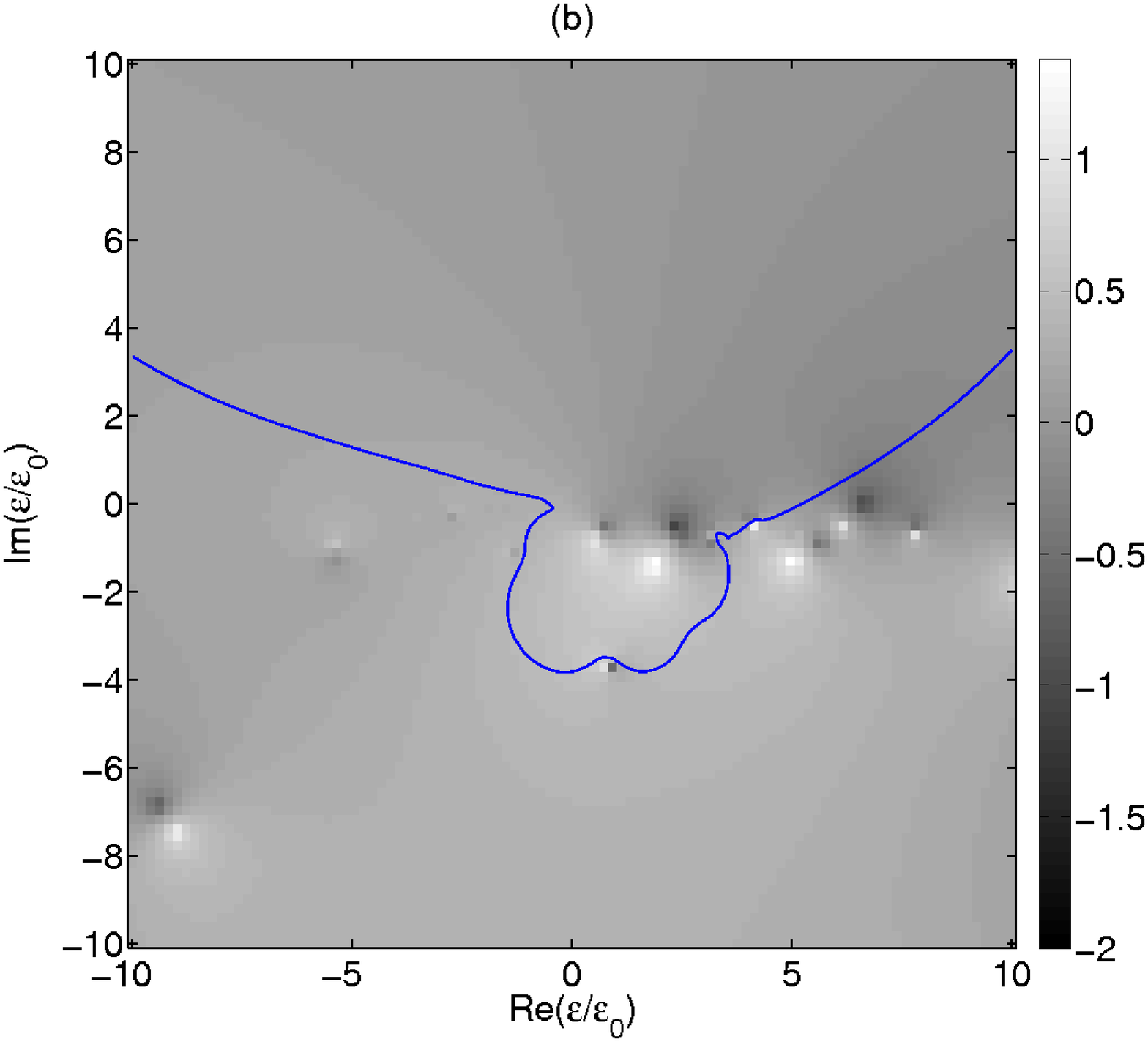,width=7cm}
\\ \vspace*{0.5cm}
\hspace*{0cm}\epsfig{file=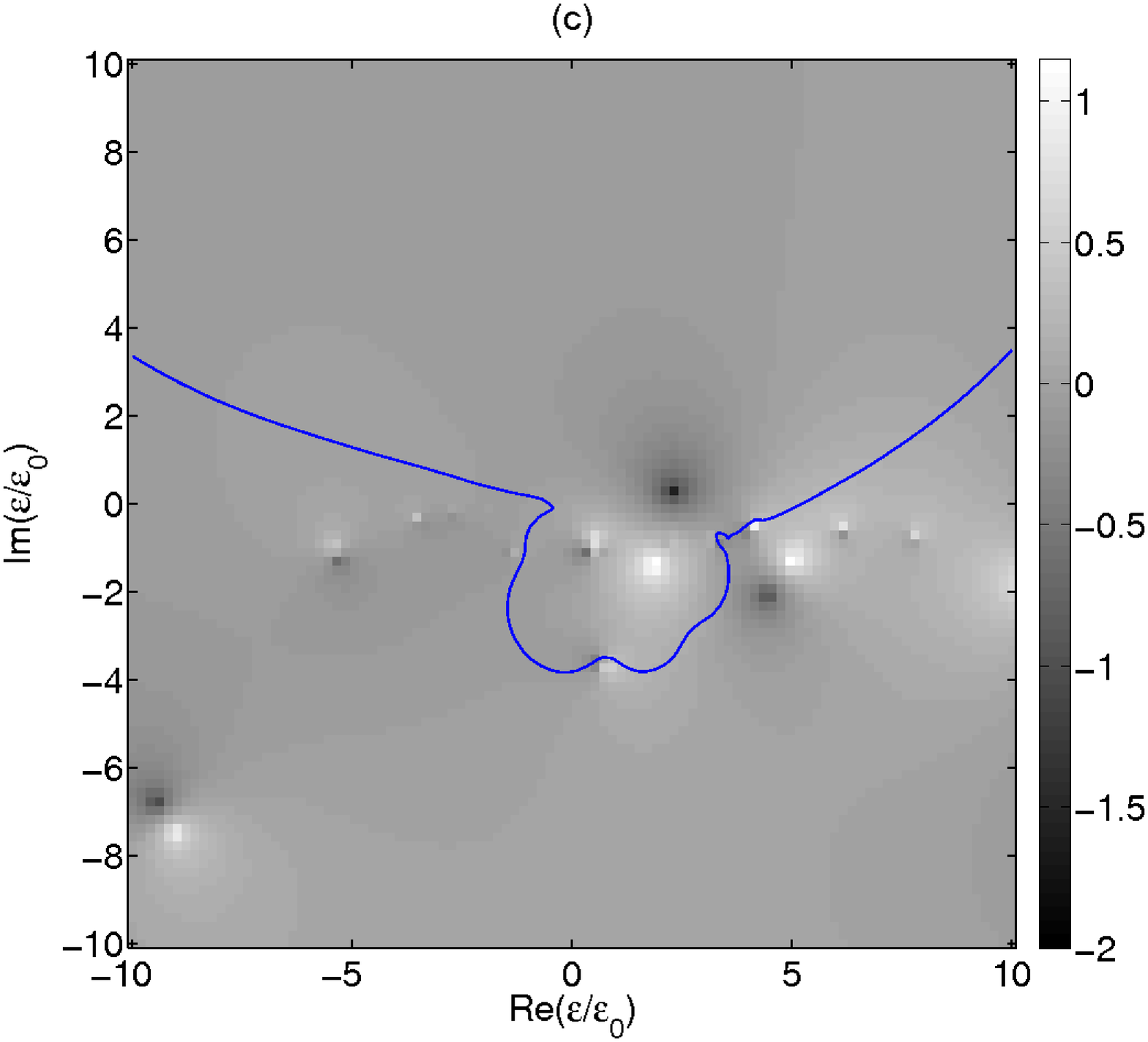,width=7cm}
\hspace*{2.5 cm}\epsfig{file=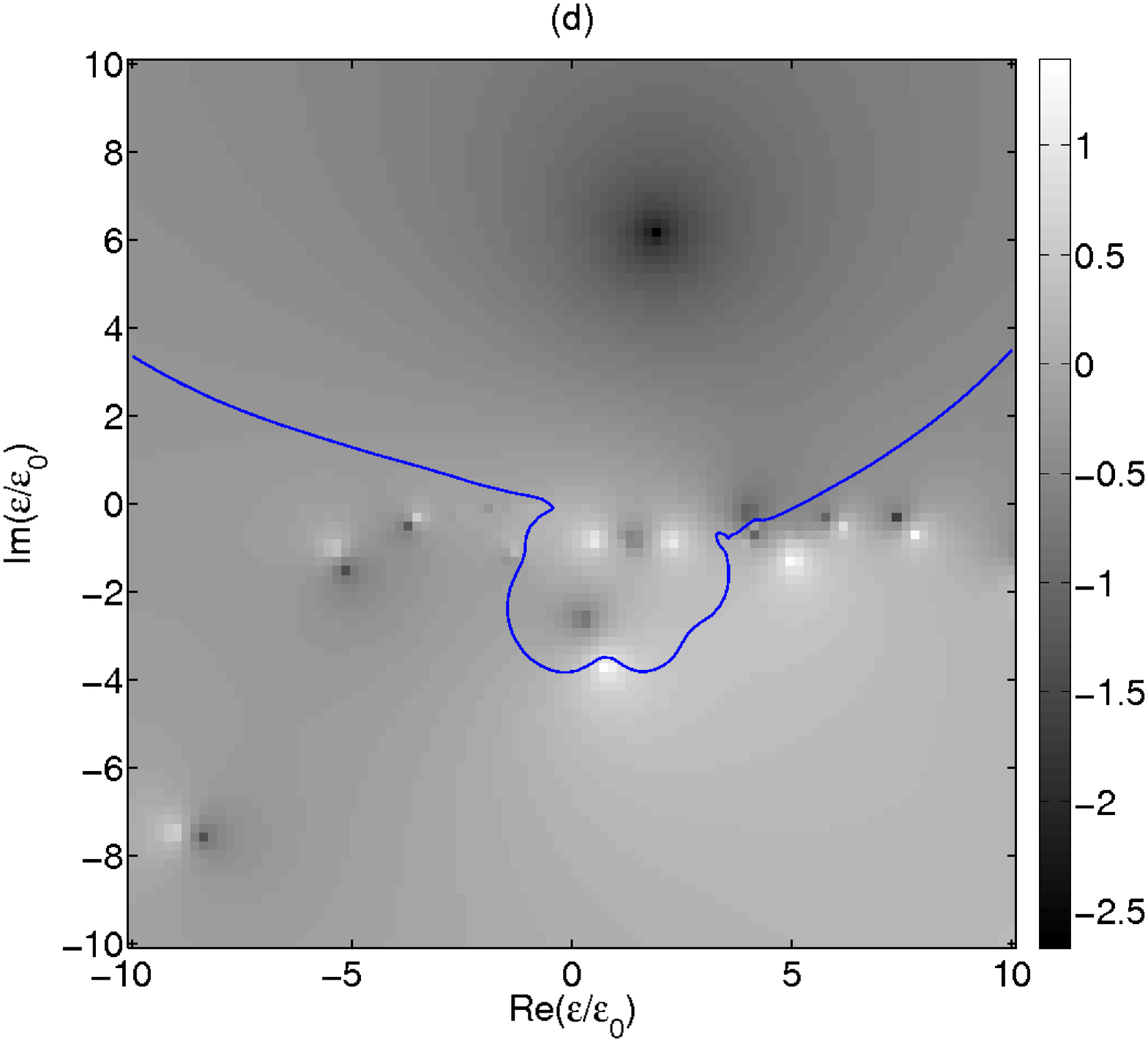,width=7cm}
\caption{An attempt to find the effective permittivity of a larger photonic-crystal sample:
(a) -- scattering configuration; (b), (c), and (d) -- data-discrepancy functionals for Receiver~1,
2, and 3, correspondingly. Although this crystal is built out of the elementary cells considered in Fig.~2,
the effective permittivities are now different as they are influenced by the geometry of the sample,
which has the same dimensions as the homogeneous object considered in Fig.~1.}
\end{figure*}
The frequency is chosen in such a way that the cyliners in Fig.~3~(a) form a triangular
photonic lattice in the horizontal cross-section with the lattice period equal to $\lambda/3$. 
Thus, we are in the vicinity of the
bandgap for this lattice, which might be the reason behind the large losses exhibited by one of the effective permittivity minima,
see Fig.~3~(d). Otherwise there are still multiple minima for each separate receiver location,
and no stationary minima. Also the minima of Fig.~2 and Fig.~3 are different, showing 
that the effective permittivity depends on the geometry and size of the sample.
Of course, it would be interesting to see, if there is a convergence in the location of the minima for progressively 
larger pieces of this photonic crystal. Unfortunately, we were not able to investigate this question, due to the 
limitations imposed on us by the computing resources. 

The Arnoldi algorithm used in \cite{Budko2004} and here is still a little too costly
in terms of computer memory as it requires storage of $M$ vectors of size $N$,
which are used as a basis for the total field. The reason for this choice of a robust but 
computationally expensive algorithm is the non-normal complex-symmetric system 
matrix of our problem. 
A possible alternative based on the Pade via Lanczos process was proposed 
in \cite{Remis2006} for the two-dimensional effective inversion problem. 
It requires storage of only three vectors of size $N$. A further generalization 
of the method, suitable for inhomogeneous scattering models and working 
with the Maxwell equations in their differential form was proposed in \cite{Druskin2007}.

We have presented here only a few examples which illustrate the main conclusions of the
theoretical part of this paper. In addition, we have performed a large number of numerical 
experiments with different types of objects, looked at different field components,
tried different types of incident fields (plane waves and Gaussian beams),
considered measurements in the far-field zone, etc. Every time we would get
images quite similar to the ones presented here, with a large number of virtually 
unpredictable minima. Except for the low frequencies, where within the considered
range of permittivities we would typically get only one minimum, which was
more or less stable with respect to the receiver location. When we considered a
larger sample of a photonic crystal (7 rows, 12--13 cylindrical rods in each) at 
a very low frequency, then the observed minimum
was in a close agreement with the effective permittivity predicted by the Maxwell-Garnet
mixing formula.

\section{Conclusions}
We have presented a generalization of the S-parameter retrieval method
to three-dimensional finite objects and arbitrary illumination and observation 
conditions. We view it as a special kind of inverse scattering problem -- an effective
inversion problem. Many, if not all, conclusions of this paper equally apply to the S-parameter retrieval 
technique, the original low-frequency effective medium theory, and even 
to the derivation of the macroscopic Maxwell equations, as we have 
shown here the mathematical equivalence of all these problems up to
the form of averaging/scattering operators.
Of course, analysis in 3D is much more complicated and implicit
than in 1D, where an explicit analytical solution for a homogeneous
slab is available. Nevertheless, straightforward application of the
spectral analysis and basic results from the inverse scattering theory
showed that the general 3D case is similar to the well-studied 1D slabs in many respects.

The ``exact'' effective permittivity, which provides the exact match between
the scattered field from an effective homogeneous object and the original 
inhomogeneous one, exists only on a limited set of incident fields and 
a limited number of observation points (angles). In fact, we can only be sure
about the existence of this exact effective permittivity, either ``physical'' or not, when we have a single
incident field and a single component of the scattered field observed
at a single spatial location. We have shown that addition of just one more observation
location may already cause the non-existence of the exact effective permittivity.
On the other hand, addition of sources and/or receivers may 
lead to the uniqueness of an approximate effective permittivity 
(although we are not able to prove it yet).
However, the more data we take into account the more approximate
(less accurate) an effective model becomes. Also, the approximate
effective permittivity is geometry-dependent, making it difficult
to view metamaterials and other strongly inhomogeneous
composites as continuous media suitable for carving out 
arbitrarily-shaped optical devices.

The exact effective permittivity, if it exists, is non-unique. 
As was shown here, this is due to the non-linearity of the
effective inversion problem and the nontrivial null-space 
of the scattered-field operator. In the single
data-point case, the number of additional exact effective permittivities
depends on the spatial spectrum of the incident field -- the broader this
spectrum, the more non-unique is the solution of the problem.
The spatial spectrum here is the spectrum of the scattering operator,
rather than the usual plane waves. An incident plane wave has, in fact,
a very broad spatial spectrum. The location of additional solutions
in the complex plane is determined by the incident field and the spectrum of the scattering 
operator and is pretty arbitrary. Hence, it is not always possible to choose 
one of the solutions on some ``physical'' grounds.

The effective inversion problem becomes singular for
certain values of effective permittivity. Many of them lie in
the ``non-physical'' area of the complex plane and show negative loss.
As such they do not cause much trouble. However, real nonpositive
values of the effective permittivity important for negative refraction and invisibility 
must be excluded as well, as for those
values the solution of the forward scattering problem either does not exist,
or is not unique, or both. Luckily, the location of singularities depends
only on the applied frequency and the outer shape of the object
and does not depend on the illumination/observation conditions.

Although we have expressed some scepticism about the usefulness of effective
models that are more complex than the original scatterer, we consider
this topic to be worth pursuing. For example, it is imperative to know
if there exists a complete-data set, perhaps, larger than the boundary data
of the Calder{\'o}n problem, such that would guarantee the uniqueness 
of the solution of the inverse scattering problem for a general inhomogeneous 
anisotropic scatterer. Also, the formal difference between the
effective inversion method and the effective medium 
(and macroscopic electrodynamics) approach,
although small, needs to be considered in more detail.
Even with respect to the completely equivalent (double-averaged) EMT we do not know 
if there are conditions ensuring the uniqueness of the averaged inverse problem (\ref{eq:Homogenization}). 
Thus, depending on the actual form of the averaging operator $B$, it may still turn out that 
the problems of the EMT and the macroscopic electrodynamics
have exact solutions under the illumination conditions which 
preclude the existence of the solution of the effective inversion (S-parameter retrieval)
problem.


\end{document}